\documentclass{article}
\usepackage{epsfig,amsmath,amssymb,times,mathtools,url,hyperref,textgreek,multirow,bm,mathtools,color,booktabs,array}
\usepackage[margin=0.7in]{geometry}
\usepackage[utf8]{inputenc}
\usepackage[english]{babel}
\DeclareMathOperator{\sign}{sgn}
\newcommand*{\matminus}{%
  \leavevmode
  \hphantom{0}%
  \llap{%
    \settowidth{\dimen0 }{$0$}%
    \resizebox{\dimen0 }{\height}{$-$}%
  }%
}
\begin{document}
\title{The Elliptical Ornstein-Uhlenbeck Process}
\author{Adam M. Sykulski\textsuperscript{1}, Sofia C. Olhede\textsuperscript{2} and Hanna M. Sykulska-Lawrence\textsuperscript{3}}
\date{}
\maketitle
%%%%%%%%%%%%%%%%%%%%%%%%%%%%%%%%%%%%%%%%%%%%%%%%%%%%%%%%%%%%%%%%%%%%%%%%%%%%%%%%%%%%%%%%%%%%%%%%%%%%%%%%%%%%%%%%%%

\begin{abstract}
We introduce the elliptical Ornstein-Uhlenbeck (OU) process, which is a generalisation of the well-known univariate OU process to bivariate time series. This process maps out elliptical stochastic oscillations over time in the complex plane, which are observed in many applications of coupled bivariate time series. The appeal of the model is that elliptical oscillations are generated using one simple first order stochastic differential equation (SDE), whereas alternative models require more complicated vectorised or higher order SDE representations. The second useful feature is that parameter estimation can be performed  semi-parametrically in the frequency domain using the Whittle Likelihood. We determine properties of the model including the conditions for stationarity, and the geometrical structure of the elliptical oscillations. We demonstrate the utility of the model by measuring periodic and elliptical properties of Earth's polar motion.
\let\thefootnote\relax\footnote{\textsuperscript{1} Department of Mathematics and Statistics, Lancaster University, UK (email: a.sykulski@lancaster.ac.uk)}
\let\thefootnote\relax\footnote{\textsuperscript{2} \'Ecole polytechnique f\'ed\'erale de Lausanne, Switzerland}
\let\thefootnote\relax\footnote{\textsuperscript{3} Department of Aeronautical and Astronautical Engineering, University of Southampton, UK}
\end{abstract}

\section{Introduction}\label{S:intro}
Complex-valued representations of bivariate time series are widely used in statistics \cite{knight2019long,wahlberg2011locally,walker1993periodogram}, signal processing \cite{schreier2010statistical,sykulski2017frequency}, and numerous application disciplines \cite{baran2018d,guillaumin2017analysis,xiong2015forecasting}. A key advantage of the complex-valued representation is that it can be conveniently used to separate structures in coupled bivariate time series that are {\em circular} or {\em noncircular} when viewed in the complex plane. In signal processing, this dichotomy is sometimes referred to as {\em proper} or {\em improper}, when specifically describing the geometry of the second-order structure of time series \cite{schreier2010statistical}. A type of noncircularity of particular interest is that of {\em elliptical oscillations} in a bivariate time series trajectory, which are observed across numerous applications including oceanography \cite{lilly2006wavelet}, seismology \cite{sykulski2016widely}, and planetary geophysics \cite{barkin2010elliptical}.

We introduce a process that can model such elliptical oscillations in continuous time. Specifically, we propose the {\em elliptical OU process} given by the following first order SDE
\begin{equation}
    dz(t) = (-\alpha_1 +i\beta_1) z(t)dt + (-\alpha_2+i\beta_2) z^*(t)dt + dW(t),
    \label{eq:WILCOU}
\end{equation}
where $z(t)=x(t)+iy(t)$, $i\equiv\sqrt{-1}$, and $z^\ast(t)$ is the complex conjugate of $z(t)$. The parameters $\{\alpha_1,\beta_1,\alpha_2,\beta_2\}$ are real-valued, and we shall place constraints on these parameters which ensure $z(t)$ is stationary in Section~\ref{S:model}. In~\eqref{eq:WILCOU}, $W(t)$ is a complex Wiener process, whose increments follow a complex normal distribution such that $B=\{W(t+\delta)-W(t)\}/\sqrt\delta \sim \mathcal{CN}(0,\sigma^2,r)$, where $\sigma^2=\mathrm{E}(BB^\ast)$ defines the variance of the complex normal, and $r=\mathrm{E}\{B^2\}$ defines the {\em pseudo-}variance and is a complex-valued quantity in general \cite{schreier2010statistical}.

If we set $\alpha_2 = \beta_2 = r = 0$ in \eqref{eq:WILCOU} then we recover the complex OU process, introduced by Arat\'o {\em et al.} \cite{arato1962evaluation}, which is a circular and proper complex-valued process. A proper process formally means that the complementary covariance defined by $r_z(\tau)=\mathrm{E}\{z(t)z(t+\tau)\}$ is zero for all $\tau$, where the autocovariance of a complex-valued process is defined by $s_z(\tau)=\mathrm{E}\{z(t)z^\ast(t+\tau)\}$. Setting $\alpha_2 = \beta_2 = r = 0$ in \eqref{eq:WILCOU}, and hence $r_z(\tau)=\mathrm{E}\{z(t)z(t+\tau)\}=0$, has the effect of ensuring the complex OU of \cite{arato1962evaluation} maps out stochastic circular oscillations with frequency $\beta_1$ and damping $\alpha_1>0$. The complex OU was proposed by \cite{arato1962evaluation} specifically to study the Chandler wobble---a small oscillatory deviation in the Earth's axis of rotation, but has also been used in numerous other physical applications including physical oceanography \cite{sykulski2016lagrangian}, magnetic fields, and reaction-diffusion systems \cite{baran2018d}.

The purpose of this paper is to study equation~\eqref{eq:WILCOU} in the more general case $\alpha_2 \neq \beta_2 \neq r \neq 0$. In Section~\ref{S:model} we derive properties including conditions for stationarity, the analytical form of the power spectral density, and the geometrical relationship between the SDE parameters and the properties of the elliptical oscillations (e.g. the eccentricity and orientation). In Section~\ref{S:estim} we provide computationally-efficient techniques for fitting parameters of our model to sampled time series, either using a fully parametric approach, or a semi-parametric approach when the model is misspecified at certain frequencies. Finally, in Section~\ref{S:App} we demonstrate the applicability of our model by studying the elliptical oscillations contained within Earth's polar motion, thus extending the earlier analyses of Arat\'o {\em et al.} \cite{arato1962evaluation} and Brillinger \cite{brillinger1973empirical} who restricted findings to capturing the properties of strictly circular oscillations.

\subsection{Relationship to Literature}
The literature on stochastic modelling of noncircular and improper complex-valued time series has primarily focused on linear filters of discrete-time processes, see e.g. \cite{navarro2008arma,picinbono1995widely,sykulski2016widely}. To create noncircularity/impropriety these filters take a {\em widely linear} form by applying
autoregressive and moving average terms to complex-valued processes and their complex conjugates, taking the general form
\begin{equation}
z_t = \sum_{j=1}^p g_j z_{t-j} + \sum_{j=1}^p h_j z^\ast_{t-j} + \sum_{j=0}^q k_j \epsilon_{t-j} + \sum_{j=0}^q l_j \epsilon^\ast_{t-j},
\label{eq:WLARMA}
\end{equation}
where $\epsilon_t$ is i.i.d complex-proper noise. In this context our generating SDE of~\eqref{eq:WILCOU} can be interpreted as the continuous-time analogue of the AR(1) version of~\eqref{eq:WLARMA} (with $p=1$, $q=0$) studied in~\cite{sykulski2016widely}. This is consistent with OU processes being considered the continuous-time analogue of AR(1) processes in general. However, as we shall see, the mapping between \eqref{eq:WILCOU} and \eqref{eq:WLARMA} is non-trivial meaning the processes are worth studying separately in their own right---as has been shown to generally be the case between CARMA (continuous-time ARMA) and discrete-time ARMA models \cite{brockwell2001continuous,chan1987note}.

In this paper we focus on continuous-time processes, as in many physical applications it is preferable to model the evolution of a time series in continuous time using SDEs, rather than discrete-time filters. This is because SDE representations allow explicit connections to be made with underlying dynamical equations (see e.g. \cite{baran2018d,veneziani2004oceanic}), and also provide a more robust modelling framework to deal with high frequency data \cite{brockwell2013high}. In the context of complex-valued time series, continuous-time models have been considered in \cite{oya2011widely} who use Karhunen-Lo{\`e}ve expansions to generate improper continuous-time nonstationary time series. Here we focus on a stationary model and go into depth in terms of understanding its statistical properties, as well as providing techniques for parameter estimation, and a demonstration of its applicability to a real-world problem.
%%%%%%%%%%%%%%%%%%%%%%%%%%%%%%%%%%%%%%
\section{Properties of the elliptical OU process}\label{S:model}

\subsection{Process Realisations}\label{ss:realisations}
\begin{figure}
    \centering
    \includegraphics[width=2.4in]{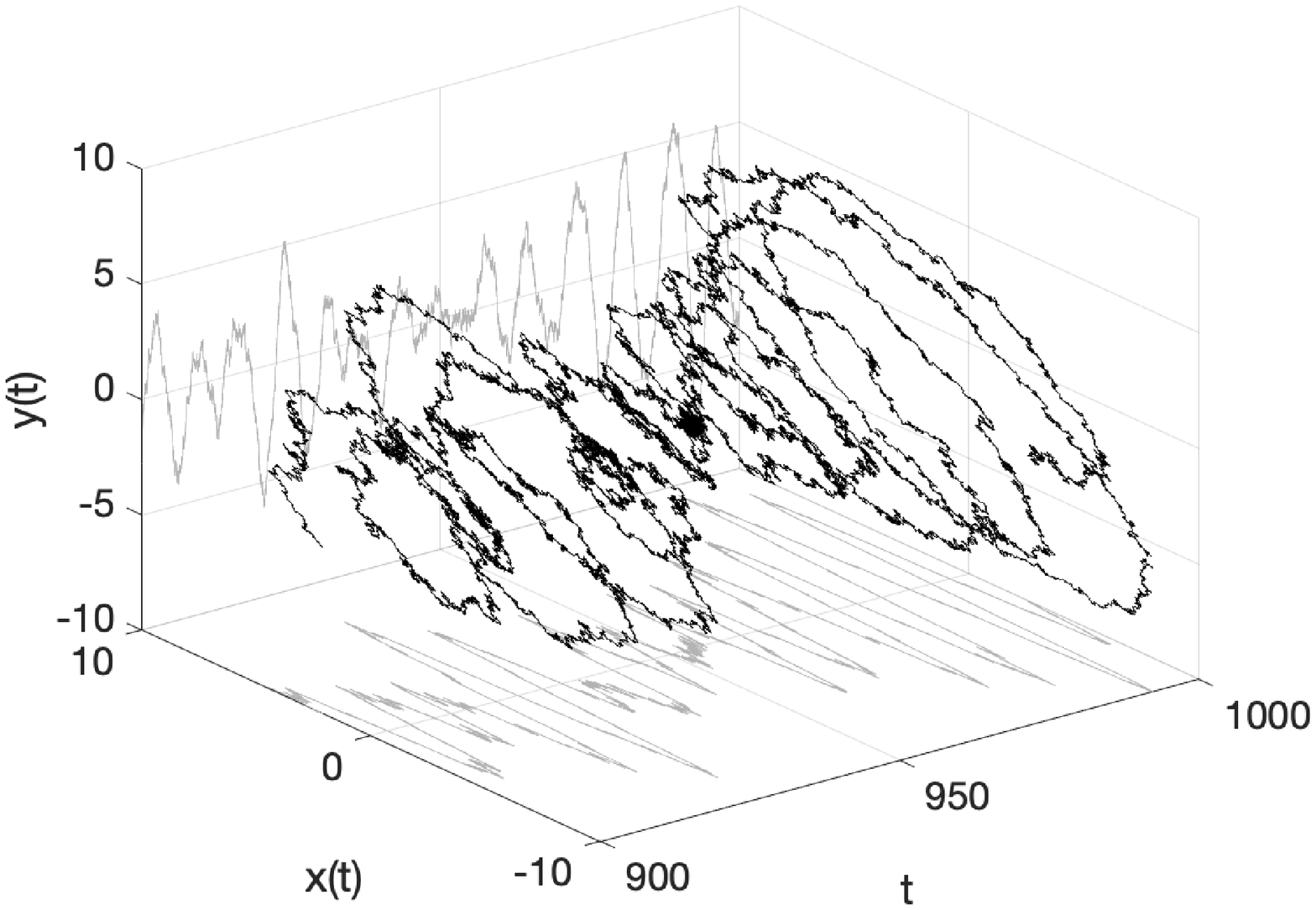} \hspace{1.5cm}
    \includegraphics[width=2.4in]{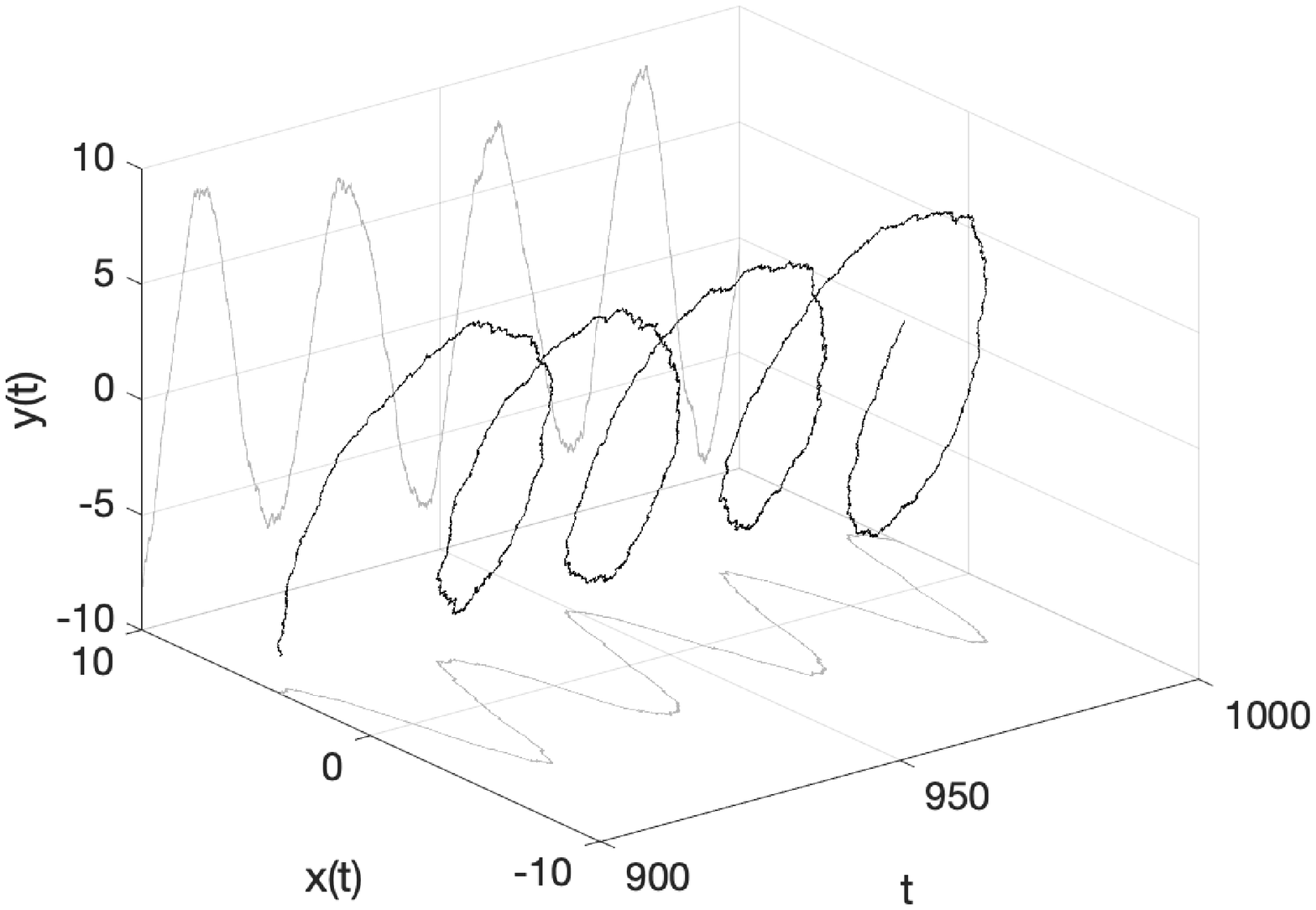} \\ \vspace{1mm}
    \includegraphics[width=2.3in]{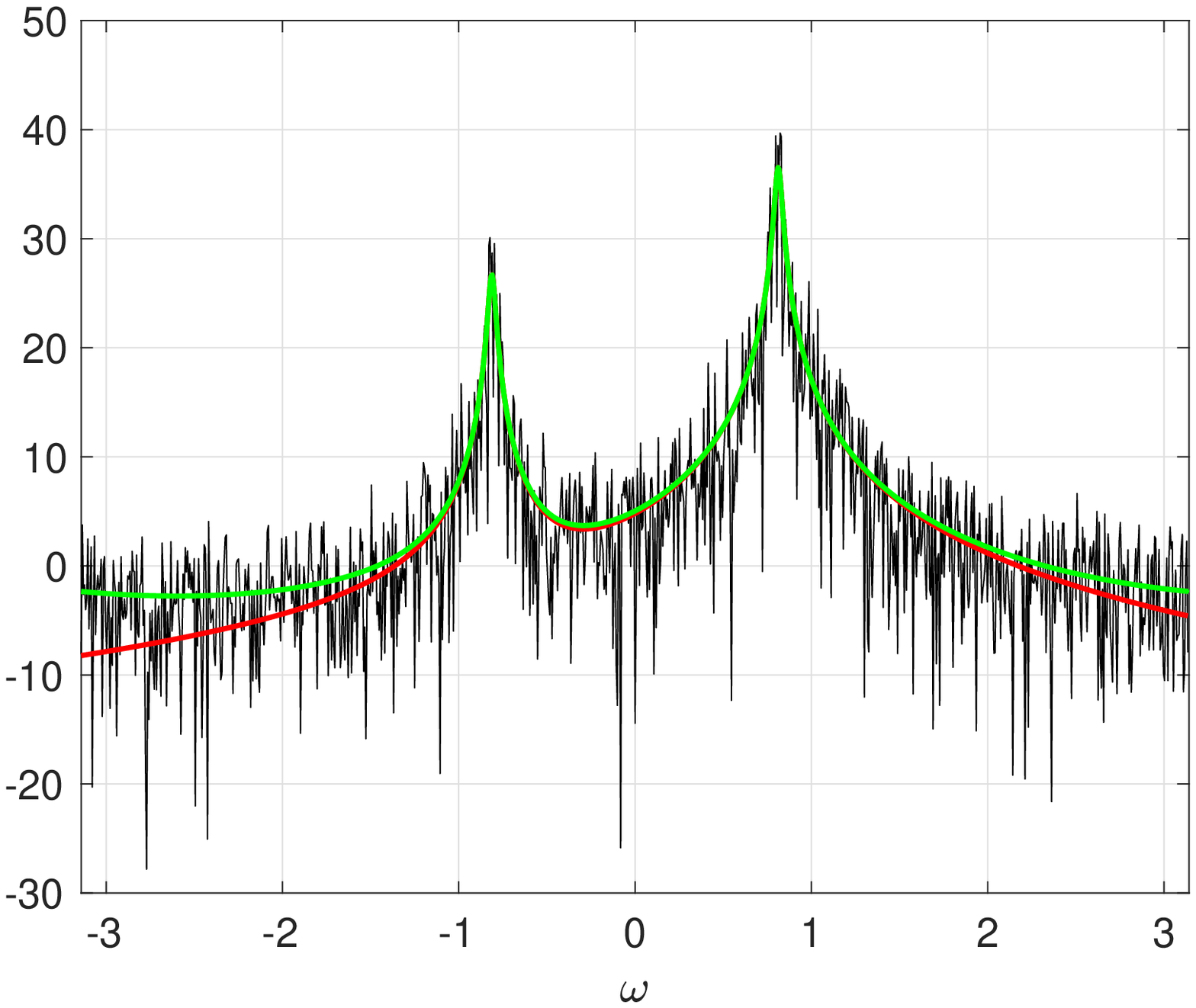} \hspace{1.5cm}
    \includegraphics[width=2.3in]{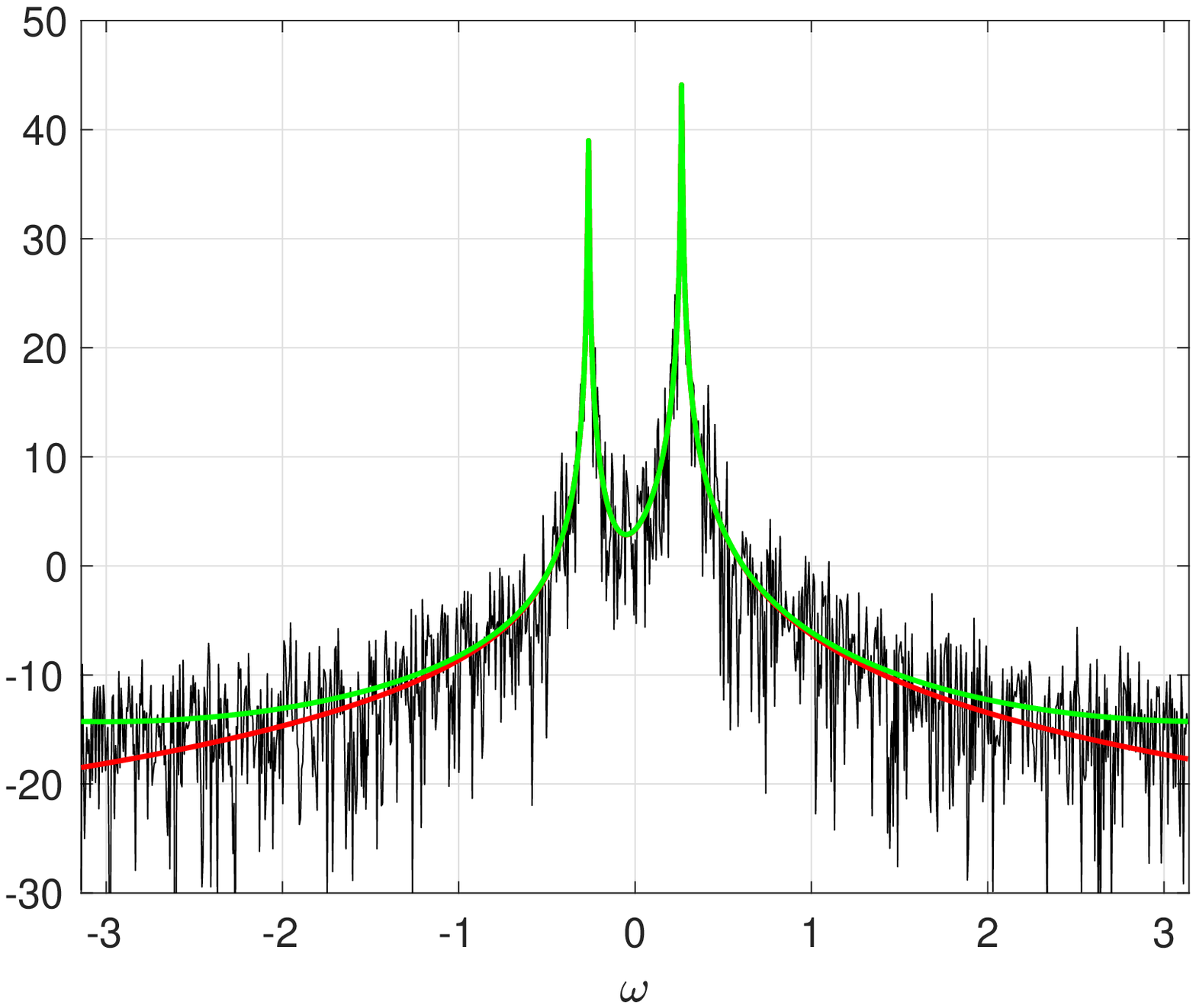}
    \caption{The top row displays two realisations of the elliptical OU process of~\eqref{eq:WILCOU} with $\{\alpha_1,\beta_1,\alpha_2,\beta_2,\sigma^2,r\}=$ $\{0.02,1,-0.5,-0.3,2,0.6+i\}$ (left) and $\{\alpha_1,\beta_1,\alpha_2,\beta_2,\sigma^2,r\}=\{0.002,0.5,0.3,0.3,0.15,-0.09-0.09i\}$ (right). The time series $z(t)$ is in black, and the $\{x(t),y(t)\}$ components are in grey. We simulate from $t=0$ to $t=1000$ but plot from $t=900$ to $t=1000$. In the bottom row we display empirical periodograms in black (on a decibel scale) of the full series sampled at integer values of $t$, and we overlay the theoretical power spectral density from~\eqref{eq:psd2} in red, with the aliased version in green.}
    \label{fig:EM}
\end{figure}

In Fig.~\ref{fig:EM} we show two realisations of the elliptical OU process under two different sets of parameter values, along with their empirical periodogram estimates to the power spectral density from sampled observations. The time series are generated using the Euler-Maruyama scheme. These realisations explain the use of the term ``elliptical'' to describe the process, as stochastic elliptical paths are being traced out over time. For complex-valued time series, the periodogram is in general asymmetric over positive and negative frequencies, as directions of spin are separated in complex-valued time series modelling. Negative frequencies correspond to clockwise oscillations and positive frequencies correspond to anti-clockwise oscillations. In each panel we overlay the theoretical power spectral density whose functional form will be derived in Section~\ref{SS:psd}, where we also include the effect of aliasing from sampling. As demonstrated in the figure, the process accomplishes generating elliptical oscillations using a simple first order model. These elliptical oscillations are seen to have differing eccentricities, orientations, and rates of damping in each example. The oscillations create two peaks of different magnitude in the power spectral density, located at the same corresponding negative and positive frequency.

Equation~\eqref{eq:WILCOU} specifies the evolution or dynamics of $z(t).$
If we try to decompose $z(t+\delta t)$ given $z(t)$, then as $dW(t)$ is not predictable, we have that
\begin{align}
\label{sde2}
    z(t+\delta t)-z(t) =&\int^{t+\delta t}_{t}
    \left\{(-\alpha_1 + i\beta_1) z(t')dt' + (-\alpha_2+i\beta_2) z^*(t')dt' + dW(t')\right\}.
\end{align}
We see directly from this equation that the increment is a widely linear transformation of $z(t)$ to produce $z(t+\delta t)$. Starting from the notion of complex geometry~\cite[Ch.~1]{lindell1992methods}, we can describe the linear vector space mapped out by complex vectors. In this space the notion of a line has been replaced by an ellipse. This ellipse can collapse to a line or a circle under special circumstances, if perturbed by $dW(t')$, as we shall shortly show.

Thus at every time point $t'$, a modification is formed by adding an ellipse to the current position $z(t)$ to get to $z(t+\delta t)$. And as $z(t)$ is fixed, if we view the process conditionally on its starting point, then~\eqref{sde2} maps out a sequence of superimposed ellipses. To understand the geometry of this ellipse we consider a deterministic version of~\eqref{eq:WILCOU} and \eqref{sde2} where $dW(t)=0$. Expressing this in terms of $x(t)$ and $y(t)$ we have that
\begin{align}
\label{eq:sde3}
dx(t) + idy(t) = & \left\{-(\alpha_1+\alpha_2)x(t)+(\beta_2-\beta_1)y(t)\right\}dt  + i \left\{(\beta_1+\beta_2)x(t)+(\alpha_2-\alpha_1)y(t) \right\}dt,
\end{align}
such that the parameter $\alpha_1$ sets the damping of the process in both $x(t)$ and $y(t)$ if it is greater than zero---as is the case with the regular real-valued OU process. The parameters $\{\alpha_2,\beta_1,\beta_2\}$ set the geometry of the ellipse of the deterministic motion as they cause asymmetric interactions between $x(t)$ and $y(t)$. As discussed already, the ellipse becomes a circle if $\alpha_2=\beta_2 = 0$, and this can be clearly seen from~\eqref{eq:sde3} when the damping $\alpha_1$ is set to zero. The other extreme is when the ellipse becomes a line which occurs when $\beta_1^2 = \alpha_2^2+\beta_2^2$. To show this is the case, consider~\eqref{eq:sde3} where linear motion occurs when $x(t)=Cy(t)$ (where $C$ is some constant). Setting the damping $\alpha_1=0$ again, from~\eqref{eq:sde3} we then solve for the simultaneous equations $-\alpha_2 = C(\beta_2-\beta_1)$ and $\beta_1+\beta_2 = C\alpha_2$ which yields $\beta_1^2 = \alpha_2^2+\beta_2^2$ for the special case of linear motion. These special cases will be verified in Section~\ref{SS:bvou} where we formally derive the eccentricity of the elliptical oscillations of the stochastic process of~\eqref{eq:WILCOU}.

This geometric structure can be related to time delay embedding plots~\cite{kantz2004nonlinear}.
In an embedding plot, $\Re\{z(t)\}$ would be plotted against $\Im\{z(t)\}$ across time, and this will capture the dynamics of the SDE as encapsulated by the ellipse geometry. Finally, note that~\eqref{sde2} is a continuous-time specification. It demonstrates that increments in the process $z(t)$ associated with arbitrary increments $\delta t$ are arrived at by a widely linear operation with some noisy offset. Equation~\eqref{sde2} also shows that $z(t)$ will trace out a continuous-time trajectory in the plane, as specified by $\{x(t),y(t)\}$.

\subsection{Process Properties}\label{SS:bvou}
To further understand the properties of~\eqref{eq:WILCOU}, we first define a {\em circular} real-valued bivariate OU process (see also \cite{veneziani2004oceanic}) given by
\begin{equation}
\begin{bmatrix} d\tilde x(t) \\ d\tilde y(t) \end{bmatrix} = 
\begin{bmatrix} \matminus\alpha & \matminus\beta \\ \beta & \matminus\alpha \end{bmatrix}
\begin{bmatrix} \tilde x(t) \\ \tilde y(t)
\end{bmatrix} dt + \frac{A}{\sqrt{2}}
\begin{bmatrix}
dW_1(t) \\ dW_2(t)
\end{bmatrix},
\label{eq:bvou}
\end{equation}
where $\alpha>0$ ensures stationarity, $\beta\in\mathbb{R}$ sets the frequency of the circular oscillation, and $dW_1(t)$ and $dW_2(t)$ are independent real-valued Wiener process increments such that $\{W_1(t+\delta)-W_1(t)\}/\sqrt\delta \sim \mathcal{N}(0,1)$ and $\{W_2(t+\delta)-W_2(t)\}/\sqrt\delta \sim \mathcal{N}(0,1)$. We refer the reader to~\cite{vatiwutipong2019alternative} for a more general overview of multivariate OU processes. Setting $\tilde z(t) = \tilde x(t) + i\tilde y(t)$ recovers the complex OU process of \cite{arato1962evaluation}. In other words, \eqref{eq:WILCOU} and \eqref{eq:bvou} are equivalent when $\alpha_1=\alpha$, $\beta_1=\beta$, $\sigma^2=A^2$, and $\alpha_2 = \beta_2 = r = 0$.

We now transform~\eqref{eq:bvou} to create elliptical oscillations by defining a new process
\begin{equation}
\begin{bmatrix}
x(t) \\ y(t)
\end{bmatrix} = QP
\begin{bmatrix}
\tilde x(t) \\ \tilde y(t)
\end{bmatrix},
\label{eq:bvou2}
\;
Q =
\begin{bmatrix}
\cos\psi & \matminus\sin\psi
\\
\sin\psi & \cos\psi
\end{bmatrix},
\;
P =
\begin{bmatrix}
\frac{1}{\rho} & 0
\\
0 & \rho
\end{bmatrix}.
\end{equation}
The parameter $\rho$ is a stretching parameter, and $\psi$ is a rotation parameter, which respectively set the eccentricity and orientation of the elliptical oscillations. For uniqueness we restrict $0<\rho\le 1$ and $-\pi/2 \le \psi \le \pi/2$. Note that $P$ must be applied first in~\eqref{eq:bvou2} for $Q$ to have an effect. We can interpret~\eqref{eq:bvou2} as a physical deformation of the circular process of~\eqref{eq:bvou}.

We now express $[x(t) \, y(t)]^T$ as a self-contained bivariate SDE by combining \eqref{eq:bvou} and \eqref{eq:bvou2} such that

\begin{equation}
\begin{bmatrix} dx(t) \\ dy(t) \end{bmatrix} = QP\left\{
\Omega P^{-1}Q^T
\begin{bmatrix} x(t) \\ y(t)
\end{bmatrix} dt + \frac{A}{\sqrt{2}}
\begin{bmatrix}
dW_1(t) \\ dW_2(t)
\end{bmatrix}\right\},
\label{eq:bvou3}
\end{equation}
where we use that $Q^{-1}=Q^T$ and where we define
\[
\Omega =\begin{bmatrix} \matminus\alpha & \matminus\beta \\ \beta & \matminus\alpha \end{bmatrix}.
\]

Equation~\eqref{eq:bvou3} is a complicated vectorised expression for generating elliptical oscillations, which we contrast with the simpler expression given in~\eqref{eq:WILCOU} using the complex representation. However,~\eqref{eq:bvou3} is useful for understanding the dynamics of, and placing parameter constraints on~\eqref{eq:WILCOU}, as we shall now show. Specifically, we set $z(t) = x(t) + iy(t)$ and show that~\eqref{eq:bvou3} can then be written in the form of~\eqref{eq:WILCOU}. To do this we define the relationship
\begin{equation}
\begin{bmatrix}
x(t) \\ y(t)
\end{bmatrix} = \frac{1}{2}T
\begin{bmatrix}
z(t) \\ z^*(t)
\end{bmatrix}, \quad T = \begin{bmatrix}
1 & 1 \\ \matminus i & i
\end{bmatrix}.
\label{eq:bvou4}
\end{equation}
By combining \eqref{eq:bvou3} and \eqref{eq:bvou4} we then obtain
\begin{equation}
    \begin{bmatrix} dz(t) \\ dz^*(t) \end{bmatrix} = \frac{1}{2} T^H L T \begin{bmatrix} z(t) \\ z^*(t) \end{bmatrix} dt + T^HQP\frac{A}{\sqrt{2}} \begin{bmatrix} dW_1(t) \\ dW_2(t) \end{bmatrix},
    \label{eq:WILCOU2}
\end{equation}
where $L = QP\Omega P^{-1}Q^T$. The elliptical OU SDE is then obtained from expanding \eqref{eq:WILCOU2} and taking the top row, such that we obtain
\begin{align}
    \label{eq:WILCOU3}
    dz(t) & =  \left(-\alpha + i\frac{\beta}{2}\left\{\frac{1}{\rho^2}+\rho^2\right\}\right)z(t) dt + \frac{\beta}{2}\left\{ \frac{1}{\rho^2} - \rho^2 \right\} (\sin 2\psi - i \cos 2\psi) z^*(t) dt + dW(t),
\end{align}
where the increment process $dW(t)$ is defined by
\[
\sigma^2 = \frac{A^2}{2}\left(\frac{1}{\rho^2}+\rho^2\right), \quad r=\frac{A^2}{2}\left(\frac{1}{\rho^2}-\rho^2\right)e^{i2\psi}. \]
By equating the parameters in~\eqref{eq:WILCOU} and \eqref{eq:WILCOU3} we can obtain an exact one-to-one mapping between the parameter set $\{\alpha_1,\beta_1,\alpha_2,\beta_2,\sigma^2\}$ of the complex SDE of \eqref{eq:WILCOU}, and the parameter set $\{\alpha,\beta,\rho,\psi,A^2\}$ of the bivariate SDE of \eqref{eq:bvou3}. The mapping in each direction is given in Table~\ref{Tab1}. The parameter $r$, which sets the pseudo-variance of the complex-valued increment process $dW(t)$, is redundant and should be set as \[
r= - \frac{\sigma^2}{\beta_1}(\beta_2+i\alpha_2),\] such that the elliptical OU process is reduced to five free parameters from mapping to an elliptically transformed bivariate OU process.
Setting $\rho=1$ in~\eqref{eq:WILCOU3} (and Table~\ref{Tab1}) recovers the three-parameter complex OU of \cite{arato1962evaluation} and~\eqref{eq:bvou}.

\begin{table*}
\begin{center}
\caption{\label{Tab1}This table provides a mapping between the parameters of the elliptical OU process of~\eqref{eq:WILCOU} and the bivariate process of~\eqref{eq:bvou3}. The function ${\tt atan2}$ is the four quadrant inverse tangent and $\sign$ is the signum function.}
\begin{tabular}{|l|l|} \hline
\multicolumn{1}{|c|}{Bivariate SDE to Elliptical OU} & \multicolumn{1}{c|}{Elliptical OU to Bivariate SDE} \\ \hline  & \\ [-2ex]
$\alpha_1 = \alpha$
&  
$\alpha = \alpha_1$ \\ [2ex]

$\beta_1 = \frac{\beta}{2}\left(\rho^2 + \frac{1}{\rho^2} \right)$
& 
$\beta  = \sign(\beta_1)\sqrt{\beta_1^2-\alpha_2^2-\beta_2^2}$\\ [2ex]

$\alpha_2 = \frac{\beta}{2}\left(\rho^2-\frac{1}{\rho^2} \right) \sin 2\psi$
& 
$\rho = \left(\frac{|\beta_1|-\sqrt{\alpha_2^2+\beta_2^2}}{|\beta_1|+\sqrt{\alpha_2^2+\beta_2^2}}\right)^{1/4}$\\ [3ex]
$\beta_2 = \frac{\beta}{2}\left(\rho^2-\frac{1}{\rho^2} \right) \cos 2\psi$ 
& 
$\psi = \frac{\sign(\matminus\beta_1)}{2} \tt{atan2}(\alpha_2,\sign(\matminus\beta_1)\beta_2)$\\ [2ex]
$\sigma^2=\frac{A^2}{2}\left(\rho^2 + \frac{1}{\rho^2} \right)$ 
 &
$A^2 = \sigma^2 \frac{\sqrt{\beta_1^2-\alpha_2^2-\beta_2^2}}{|\beta_1|}$ \\ [2ex] \hline
\end{tabular}
\end{center}
\end{table*}

In the more general setting, we observe from Table~\ref{Tab1} the simple relationship that $\alpha_1=\alpha$, meaning $\alpha_1$ sets the damping rate of the oscillations in~\eqref{eq:WILCOU}, and we thus require $\alpha_1>0$ for the elliptical OU process to be stationary. The parameters $\{\beta_1,\alpha_2,\beta_1\}$ jointly determine $\{\beta, \rho, \psi\}$ (the oscillation frequency, eccentricity and orientation) and we require $|\beta_1| > \sqrt{\alpha_2^2 + \beta_2^2}$ to create a valid mapping between the two processes. The eccentricity of the oscillations is given by
\[
\varepsilon = \sqrt{1-\rho^4} = \sqrt{\frac{2\sqrt{\alpha_2^2+\beta_2^2}}{|\beta_1|+\sqrt{\alpha_2^2+\beta_2^2}}},
\]
such that larger values of $\alpha_2$ and $\beta_2$ create more eccentric oscillations. This formally establishes the geometric properties of the elliptical oscillations and verifies the results from Section~\ref{ss:realisations} that the oscillations are circular when $\alpha_2=\beta_2=0$, and collapse to a line as $\alpha_2^2+\beta_2^2$ approaches $\beta_1^2$. In the next section we derive the power spectral density of the elliptical OU process which will provide yet further intuition on the effect of the different parameters.

Overall, we see that complex-valued modelling provides a much more straightforward SDE representation of elliptical oscillations than bivariate modelling, as shown in~\eqref{eq:WILCOU} and Fig.~\ref{fig:EM}. However, mapping to an underpinning bivariate process, as in~\eqref{eq:bvou3}, allows us to further understand the geometry and dynamics of the elliptical oscillations, as well as place necessary parameter constraints.

A similar mapping analysis was performed with discrete-time models in~\cite{sykulski2016widely} by equating a widely linear autoregressive AR(1) process to a corresponding bivariate AR(1) process. The mappings between the parameters are significantly different here as compared with those found in~\cite{sykulski2016widely} for discrete time. There are two reasons why these mappings are so different. First, although an AR(1) process can generally be interpreted as a discrete-time analogue of an OU process, there is no simple transformation between their sets of parameters in the widely linear case, as we show in Appendix~A. This is consistent with~\cite{brockwell2001continuous,chan1987note} who discuss the nontrivial relationship between sampled CARMA (continuous-time ARMA) models and regular discrete-time ARMA models. Secondly, the elliptical OU of~\eqref{eq:WILCOU} has coefficients given in Cartesian form, whereas the coefficients of the widely linear AR(1) are given in polar form (see~\eqref{eq:WILCAR} in Appendix~A). These parameterizations in each case make sense as the mappings between the OU and AR processes are then straightforward in the regular (non widely linear) case, see \eqref{eq:appa1} in Appendix~A. However, these choices of parameterizations cause further departures in the parameter mappings in the widely linear case. As a result, the conditions for stationarity, and the geometrical properties of elliptical oscillations, are entirely different in the continuous-time elliptical OU proposed in this paper, and the discrete-time AR(1) proposed in~\cite{sykulski2016widely}.
%%%%%%%%%%%%%%%%%%%%%%%%%%%%%
\subsection{The Power Spectral Density}\label{SS:psd}
For stationary complex-valued processes the power spectral density can in general be defined from the autocovariance sequence of the process, such that
\begin{equation}
\label{eqn:sdf}
S_z(\omega) = \int s_z(\tau) e^{-i\omega\tau}d\tau, \quad s_z(\tau) = \mathrm{E}\{z(t)z^*(t+\tau)\},
\end{equation}
where the frequency $\omega$ will always be given in radians in this paper. The power spectral density of the complex OU of~\cite{arato1962evaluation} is given by~\cite{sykulski2016lagrangian}
\begin{equation}
    S_{\tilde z}(\omega) = \frac{\sigma^2}{\alpha_1^2+(\omega-\beta_2)^2} = \frac{A^2}{\alpha^2+(\omega-\beta)^2},
    \label{eq:psd}
\end{equation}
which we have provided both in terms of the parameterization of~\eqref{eq:WILCOU} (with $\alpha_2=\beta_2=r=0$), and of the circular bivariate process of~\eqref{eq:bvou}. Note that despite being a proper process, the spectral density will contain energy at both negative and positive frequencies, decaying at rate $\omega^{-2}$ from the peak frequency.

The power spectral density of the elliptical OU process is given by
\begin{equation}
    S_{z}(\omega) = A^2\left\{ \frac{\left(\frac{1}{\rho}+\rho\right)^2}{\alpha^2+(\omega-\beta)^2} + \frac{\left(\frac{1}{\rho}-\rho\right)^2}{\alpha^2+(\omega+\beta)^2} \right\},
    \label{eq:psd2}
\end{equation}
which is given in terms of the parameterization of the elliptical bivariate process of~\eqref{eq:bvou2}. Then to find the power spectral density in terms of~\eqref{eq:WILCOU} one simply substitutes using the transformations in the right column of Table~\ref{Tab1}. The derivation of~\eqref{eq:psd2} is provided in Appendix~B.

Intuition is gained by examining~\eqref{eq:psd2}. While~\eqref{eq:psd} has just one peak in the spectral density located at $\omega=\beta$, \eqref{eq:psd2} has two peaks located at $\omega=\pm\beta$. The rate of damping of both peaks is determined by $\alpha$, and the ratio of magnitudes of the two peaks is determined by $\rho$. Note that the orientation parameter $\psi$ does not feature in the power spectral density. When~\eqref{eq:psd2} is represented using the parameters of~\eqref{eq:WILCOU} then we see that $\alpha_1$ defines the damping of the two peaks, and $\{\beta_1,\alpha_2,\beta_2\}$ together determine the peak locations and their relative magnitudes. We have overlaid the power spectral density of~\eqref{eq:psd2} over the periodogram of the simulated series in Fig.~\ref{fig:EM}, where we have also included the aliased spectral density given by $\tilde{S}_z(\omega) = \sum_{k=-\infty}^\infty S_z(\omega+2\pi k/\Delta)$ which accounts for the departures at high frequency between the periodogram and spectral density, caused by sampling the time series at regular intervals of $\Delta$.

To fully specify the properties of the elliptical OU process, we need to derive the {\em complementary} spectrum defined by
\[
R_z(\omega) = \int r_z(\tau) e^{-i\omega\tau}d\tau, \quad r_z(\tau) = \mathrm{E}\{z(t)z(t+\tau)\}.
\]
The complex OU of \cite{arato1962evaluation} is a  proper process and therefore $R_z(\omega)=r_z(\tau) = 0$. The elliptical OU process has a complementary spectrum given by
\begin{align}
    \label{eq:psd3}
    R_z(\omega) & = \frac{A^2}{4}\left(\frac{1}{\rho^2}-\rho^2\right) 
    \left\{ \frac{1}{\alpha^2+(\omega-\beta)^2} + \frac{1}{\alpha^2+(\omega+\beta)^2} \right\}e^{i2\psi},
\end{align}
which {\em is} dependent on $\psi$, as well as all the other parameters. We see that as long as $\rho < 1$ then $R_z(\omega)$ is non-zero such that $r_z(\tau)$ is also non-zero and the elliptical OU is an improper process as intended. The derivation of~\eqref{eq:psd3} can also be found in Appendix~B. We note that full specification of the power spectral density and complementary spectrum allows for an exact method of simulating the process at a fixed sampling rate, based on circulant embedding and Fourier transforms, as an alternative to Euler-Maruyama, see~\cite{sykulski2016exact} for details.

%%%%%%%%%%%%%%%%%%%%%%%%%%%%%%%%%%%%%%%%%%%%
\section{Parameter estimation}\label{S:estim}
The elliptical OU of~\eqref{eq:WILCOU} is an improper process, as we have shown. Therefore to estimate parameters using a maximum likelihood approach from an observed complex-valued time series, we would need to invert large matrices containing both autocovariance and complementary covariance terms, which will be computationally intensive for large sample sizes. In this section we detail how inference for complex-valued time series can be done in the frequency domain using the Whittle Likelihood and computationally-efficient Fourier transforms, see~\cite{sykulski2017frequency} for a comprehensive review.

The Whittle likelihood is a pseudo-maximum likelihood approach which has been shown, for large classes of processes, to converge at the optimal $\mathcal{O}(1/\sqrt{n})$ rate to the true parameter values as the sample size $n$ increases~\cite{dzhaparidze1983spectrum}. The classical assumptions made on the process require boundedness (from above and below) and twice differentiability of the spectral density with respect to $\omega$ (the frequency) and $\boldsymbol\theta \in \Theta$ (the parameter vector), as well as the true parameters lying in the interior of the parameter space. The elliptical OU process indeed has a spectral density that is bounded from above and below, and is twice differentiable (with $\omega$ and $\boldsymbol\theta \in \Theta$), therefore as long as the true parameters do not lie on the boundary (e.g. linear motion when $\rho = 0$) then we can expect Whittle likelihood to perform well for the elliptical OU process with large enough samples.

Consider a length-$n$ observed complex-valued time series $\mathbf{Z} = [Z_1,\ldots,Z_n]$ where the time series is regularly sampled at intervals denoted by $\Delta$. To perform inference we need two objects, the first is the Discrete Fourier Transform of the data $\mathbf{Z}$ and its conjugate $\mathbf{Z}^\ast$, denoted $J_C(\omega)$ and given by
\[
J_C(\omega) = \begin{bmatrix} J_Z(\omega) \\ J_{Z^\ast}(\omega) \end{bmatrix} = \sqrt{\frac{\Delta}{N}} \sum_{t=1}^n \begin{bmatrix} Z_t \\ Z^\ast_t \end{bmatrix} e^{-i\omega t \Delta},
\]
and the second is the spectral matrix of the model family, denoted $\boldsymbol S_C(\omega;\boldsymbol\theta)$, and given by
\[
\boldsymbol S_C(\omega;\boldsymbol\theta) = \begin{bmatrix} S_z(\omega;\boldsymbol\theta) & R_z(\omega;\boldsymbol\theta) \\ R^\ast_z(\omega;\boldsymbol\theta) & S_z(-\omega;\boldsymbol\theta)
\end{bmatrix},
\]
where for the elliptical OU process, $S_z(\omega;\boldsymbol\theta)$ and $R_z(\omega;\boldsymbol\theta)$ are as defined in~\eqref{eq:psd2} and \eqref{eq:psd3} respectively. The effect of aliasing can be incorporated by using $\tilde{S}_z(\omega;\boldsymbol \theta) = \sum_{k=-K}^K S_z(\omega+2\pi k/\Delta;\boldsymbol \theta)$ and $\tilde{R}_z(\omega;\boldsymbol \theta) = \sum_{k=-K}^K R_z(\omega+2\pi k/\Delta;\boldsymbol \theta)$ in place of $S_z(\omega;\boldsymbol\theta)$ and $R_z(\omega;\boldsymbol\theta)$ respectively, where the computation becomes exact as $K\rightarrow\infty$, but due to the relatively fast $\omega^{-2}$ decay in frequency, setting a value of $K=10$ was found to have practically converged.

To obtain parameter estimates we maximise the following pseudo-likelihood over the parameter space $\boldsymbol \theta \in \boldsymbol \Theta$,
\begin{equation}
\ell_W(\boldsymbol \theta) = - \frac{1}{2}\sum_{\omega\in\Omega} \left\{ \log |\boldsymbol S_C(\omega;\boldsymbol{\theta})| + J^\mathsf{H}_C(\omega)\boldsymbol S^{-1}_C(\omega;\boldsymbol{\theta})J_C(\omega) \right\},
\label{eq:Whittle1}
\end{equation}
where $\mathsf{H}$ denotes the Hermitian transpose, and $\Omega$ is the set of Fourier frequencies given by
\begin{equation}
\Omega =\frac{2\pi}{n\Delta}\left(-\lceil {n}/{2} \rceil +1,\ldots,-1,0,1,\ldots, \lfloor {n}/{2} \rfloor\right).
\label{fourier_frequencies}
\end{equation}
The maximisation of $\ell_W(\boldsymbol \theta)$ is performed over the parameter vector $\boldsymbol{\theta} = \{\alpha,\beta,\rho,\psi,A^2\}$, and the parameter estimates corresponding to~\eqref{eq:WILCOU} are then found using the right column of Table~\ref{Tab1}. This approach can be adapted to irregularly spaced observations using the techniques described in \cite{matsuda2009fourier}.

A semi-parametric alternative is to fit a simpler pseudo-likelihood to only the power spectral density as given by
\begin{equation}
    \ell_S(\boldsymbol{\theta}) = - \sum_{\omega\in\Omega} \left\{ \log S_z(\omega;\boldsymbol{\theta}) + \frac{I_Z(\omega)}{S_z(\omega;\boldsymbol{\theta})} \right\},
    \label{eq:Whittle2}
\end{equation}
where $I_Z(\omega)=|J_Z(\omega)|^2$ is the periodogram, and the complementary spectrum $R_z(\omega;\boldsymbol\theta)$ is not used in the fit. The aliased spectral density $\tilde{S}_z(\omega;\boldsymbol \theta)$ can be used in place of $S_z(\omega;\boldsymbol \theta)$. This semi-parametric approach has some advantages in terms of robustness to model misspecification and smaller sample sizes, as we shall shortly discuss. This parametric fit, however, can only be performed over the parameter vector $\boldsymbol{\theta} = \{\alpha,\beta,\rho,A^2\}$ as $\psi$ is not present in the power spectral density of~\eqref{eq:psd2}. To estimate $\psi$ we observe from \eqref{eq:psd3} that $\arg \{R_z(\omega)\}=2\psi$, from which we can derive the following non-parametric estimate
\begin{equation}
\hat\psi = \frac{1}{2}\left[\arg\{J_Z(\omega_\mathrm{max})\}+\arg\{J_Z(-\omega_\mathrm{max})\}\right],
\label{eq:orient}
\end{equation}
the full derivation of which is given in Appendix~C, where $\omega_\mathrm{max}$ refers to the location of the peak in the spectral density, which can be approximated from the periodogram if unknown. As with \eqref{eq:Whittle1}, the parameter estimates from \eqref{eq:Whittle2} and \eqref{eq:orient} can be expressed in the form of the parameters of~\eqref{eq:WILCOU} using the right column of Table~\ref{Tab1}.

Both likelihood procedures \eqref{eq:Whittle1} and \eqref{eq:Whittle2} can be made further semi-parametric by only including a subset of frequencies from~\eqref{fourier_frequencies} in the respective summations (see also~\cite{robinson1995gaussian}). This is useful when the Fourier transform is contaminated by high frequency noise (and high frequencies should be excluded from the fit), or the chosen model is known to only be correct in a narrow range of frequencies, perhaps because an aggregation of effects has been observed. Indeed we shall employ such procedures in Section~\ref{S:App} to separate the annual and Chandler wobble oscillations of Earth's polar motion.

Other modifications to Whittle likelihood including tapering and differencing the time series, or debiasing the estimates to account for spectral blurring (see \cite{sykulski2019debiased} for a review). We did not find such modifications to be needed for the elliptical OU process. This is because the process has a relatively small dynamic range, owing to the $\omega^{-2}$ decay in \eqref{eq:psd2}, such that there is only a small amount of spectral blurring present in the periodogram.

\subsection{Simulation Study}\label{ss:sim}

We now perform a simulation study to compare the bias and root-mean-square error (RMSE) of the parameter estimates of the elliptical OU process over 1,000 Monte Carlo replicated time series of the model presented in the left panel of Figure~\ref{fig:EM}. These parameter values were chosen as they are of similar magnitude to those observed in our application to Earth's polar motion in the next section. We compare four Whittle likelihood approaches: the full parametric likelihood of~\eqref{eq:Whittle1}, the marginal semi-parametric likelihood of~\eqref{eq:Whittle2} (combined with the non-parametric estimate of $\psi$ in \eqref{eq:orient}), and semi-parametric versions of each where only a narrowband of 49 Fourier frequencies located around each of the positive and negative peak frequencies are used ($\omega\in\pm [0.725,0.897]$). The sample size is set to match the application in the next section with $n=1759$. Therefore the semi-parametric approach uses only $98/1759 \approx 5\%$ of the Fourier frequencies available in the Fourier transform, but at the frequencies where most information about the process is contained. For the first two methods we use the approximated aliased spectral density with $K=10$.

The results are shown in Table~\ref{Tab2}. We display bias and RMSE (relative to the true value, expressed as a percentage) for each of the parameters of the elliptical OU process: $\{\alpha_1,\beta_1,\alpha_2,\beta_2,\sigma^2\}$. The two approaches, \eqref{eq:Whittle1} and \eqref{eq:Whittle2},  perform very similarly (RMSE in the range of 0.5-20\%), though the full likelihood of \eqref{eq:Whittle1} does slightly better in estimating $\{\beta_1,\alpha_2,\beta_2\}$, which are the parameters that define the shape of the ellipse---this is as expected as more information content in the complementary spectrum has been used to fit the parameters. However, the most challenging parameters to estimate are the damping and amplitude $\{\alpha_1,\sigma^2\}$ (RMSE 5-20\%) and both approaches perform similarly here. As we reduce the range of frequencies used (third and fourth rows), then both approaches have slightly worse fits as expected, though the increase in RMSE is mainly observed in the amplitude parameter $\{\sigma^2\}$ (RMSE 20-25\%) as this is the only parameter that lives at all frequencies---information about $\{\alpha_1,\beta_1,\alpha_2,\beta_2\}$ (which define the damping, frequency, orientation and eccentricity of the oscillations) live at frequencies in and around the peak frequency, which is why the semi-parametric fits generally perform very well.

\begin{figure}[h]
\centering
\includegraphics[width=3.3in]{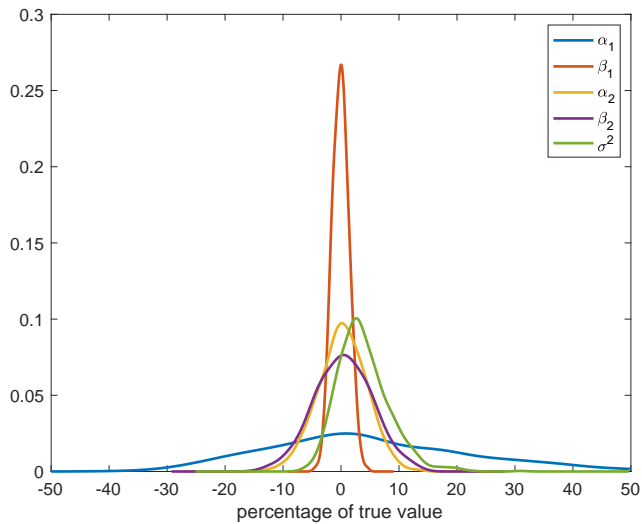}
\caption{Kernel density estimates of the deviation of parameter estimates from Table 2 using \eqref{eq:Whittle2} and $\omega\in[-\pi,\pi]$. Deviations expressed as a percentage of the true value.}
\label{figMC}
\end{figure}

In Figure~\ref{figMC} we display the kernel density estimate plots of the deviation of the parameter estimates for each parameter (relative to its true value) using~\eqref{eq:Whittle2} and $\omega\in[-\pi,\pi]$ (i.e. the second row of Table~\ref{Tab2}). The parameter estimates are approximately Gaussian, and often within 10\% of the true parameter value (except for $\alpha_1)$, suggesting the inference technique is robust, at least for the parameters and sample size considered.

The motivation behind proposing two estimation techniques, \eqref{eq:Whittle1} and \eqref{eq:Whittle2}, is because a) two distinct methods provide a validation tool for numerical optimisers that may sometimes converge differently, thus flagging parameter estimates which may be stuck at boundaries or local optima; and b) we found that both methods will (inevitably) breakdown and struggle to identify all parameters when focusing on a particularly narrowband signal. In general we have found optimising using \eqref{eq:Whittle2} to be more robust in such settings as only four parameters are fitted rather than five.

\begin{table}[h]
\begin{center}
\caption{\label{Tab2}Bias and RMSE (expressed as a percentage of the true parameter value) for 4 different Whittle inference methods of the elliptical OU process, computed using 1,000 Monte Carlo replicated time series of length $n=1759$}
\begin{tabular}{|c|c|cc|cc|cc|cc|cc|} \hline
& & \multicolumn{2}{c|}{$\alpha_1=0.02$} & \multicolumn{2}{c|}{$\beta_1=1$} & \multicolumn{2}{c|}{$\alpha_2=-0.5$} & \multicolumn{2}{c|}{$\alpha_2=-0.3$} & \multicolumn{2}{c|}{$\sigma^2=2$} \\
Method & Frequencies & Bias & RMSE & Bias & RMSE & Bias & RMSE & Bias & RMSE & Bias & RMSE \\ \hline
eq. \eqref{eq:Whittle1} & $\omega\in[-\pi,\pi]$ & 5.40 & 19.17 & 0.01 & 0.52 & -0.04 & 1.00 & 0.01 & 1.42 & 3.88 & 5.55\\
eq. \eqref{eq:Whittle1} & $\omega\in[-\pi,\pi]$ & 5.15 & 19.09 & -0.08 & 1.42 & 0.25 & 4.15 & 0.41 & 4.98 & 3.86 & 5.92\\
eq. \eqref{eq:Whittle2} & $\omega\in\pm [0.725,0.897]$ & 18.01 & 30.02 & 0.01 & 0.61 & -0.04 & 1.13 & -0.02 & 1.55 & 19.02 & 24.31\\
eq. \eqref{eq:Whittle2} & $\omega\in\pm [0.725,0.897]$ & 4.68 & 26.66 & 0.08 & 0.70 & -0.24 & 1.82 & -0.09 & 3.44 & 3.69 & 22.61\\ \hline
\end{tabular}
\end{center}
\end{table}

\subsection{Parameter Standard Errors}\label{ss:bootstrap}
Parameter standard errors and confidence intervals can be obtained from fitting a single time series using bootstrap resampling procedures developed in~\cite{dahlhaus1996frequency}. This simple procedure, developed for Whittle likelihood estimates, creates bootstrapped periodograms by multiplying a spectral density estimate by standard independent exponentially distributed random variables at each Fourier frequency, and then re-estimating parameters with each bootstrapped periodogram to obtain parameter standard errors and confidence intervals.

In Table~\ref{Tab3} we report the parameter standard errors from the Monte Carlo simulation of Table~\ref{Tab2} from \eqref{eq:Whittle2} and $\omega\in[-\pi,\pi]$. We also provide the average bootstrapped parameter standard errors using two spectral density estimators: the raw periodogram and a smoothed periodogram using the Epane\v{c}nikov kernel (bandwidth = 0.07 radians). The number of bootstrap replicates per time series is set to 100. The results are displayed in Table~\ref{Tab3}. Both bootstrap approaches provide good estimates of the standard errors. In the application section we will use the periodogram-based estimator as it is more conservative (caused by the higher variability in the periodogram yielding larger bootstrap variability).

\begin{table}[h]
\begin{center}
\caption{\label{Tab3}Parameter standard errors (as a percentage of the true value), using the same simulation setup as Table~\ref{Tab2}}
\begin{tabular}{|c|ccccc|} \hline
Technique & $\alpha_1$ & $\beta_1$ & $\alpha_2$ & $\beta_2$ & $\sigma^2$ \\ \hline
Monte Carlo & 17.64 & 1.36 &   3.89 &    4.82 &   4.04 \\
Bootstrap: Periodogram & 17.42 &    2.69 &    7.77 &  7.79 &   3.84 \\
Bootstrap: Epane\v{c}nikov & 11.78 &    1.99 &    5.68 &   5.69 &    2.81 \\ \hline
\end{tabular}
\end{center}
\end{table}
%%%%%%%%%%%%%%%%%%%%%%%%%%%%%%%%%%%%%%%%%%%%%%%%%%%%%%%%%
%%%%%%%%%%%%%%%%%%%%%%%%%%%%%%%%%%%%%%%%%%%%%%%%%%%%%%%%%
\section{Earth's polar motion}\label{S:App}
\begin{figure}[h]
\centering
\includegraphics[width=3.3in]{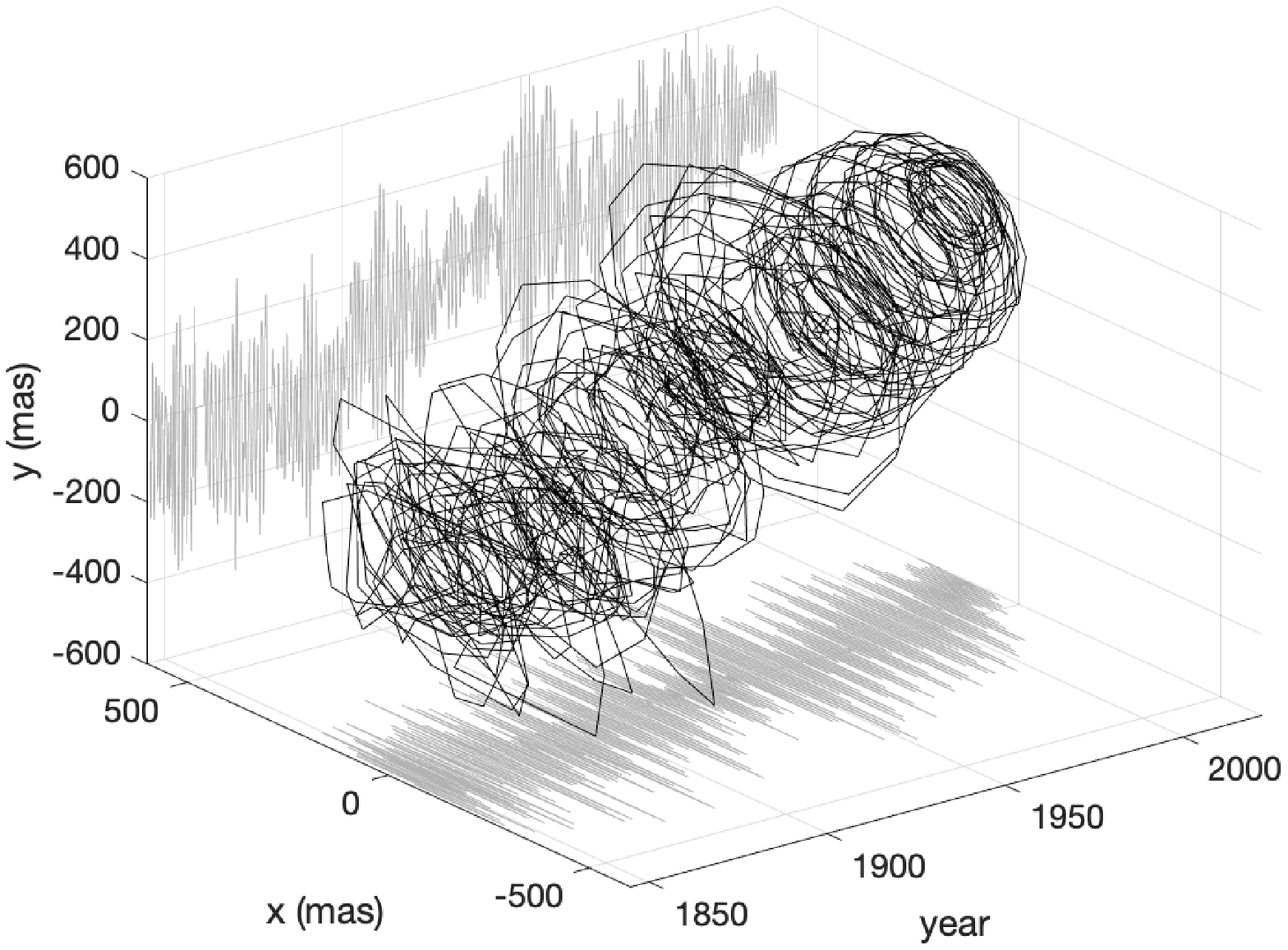}
\includegraphics[width=3in]{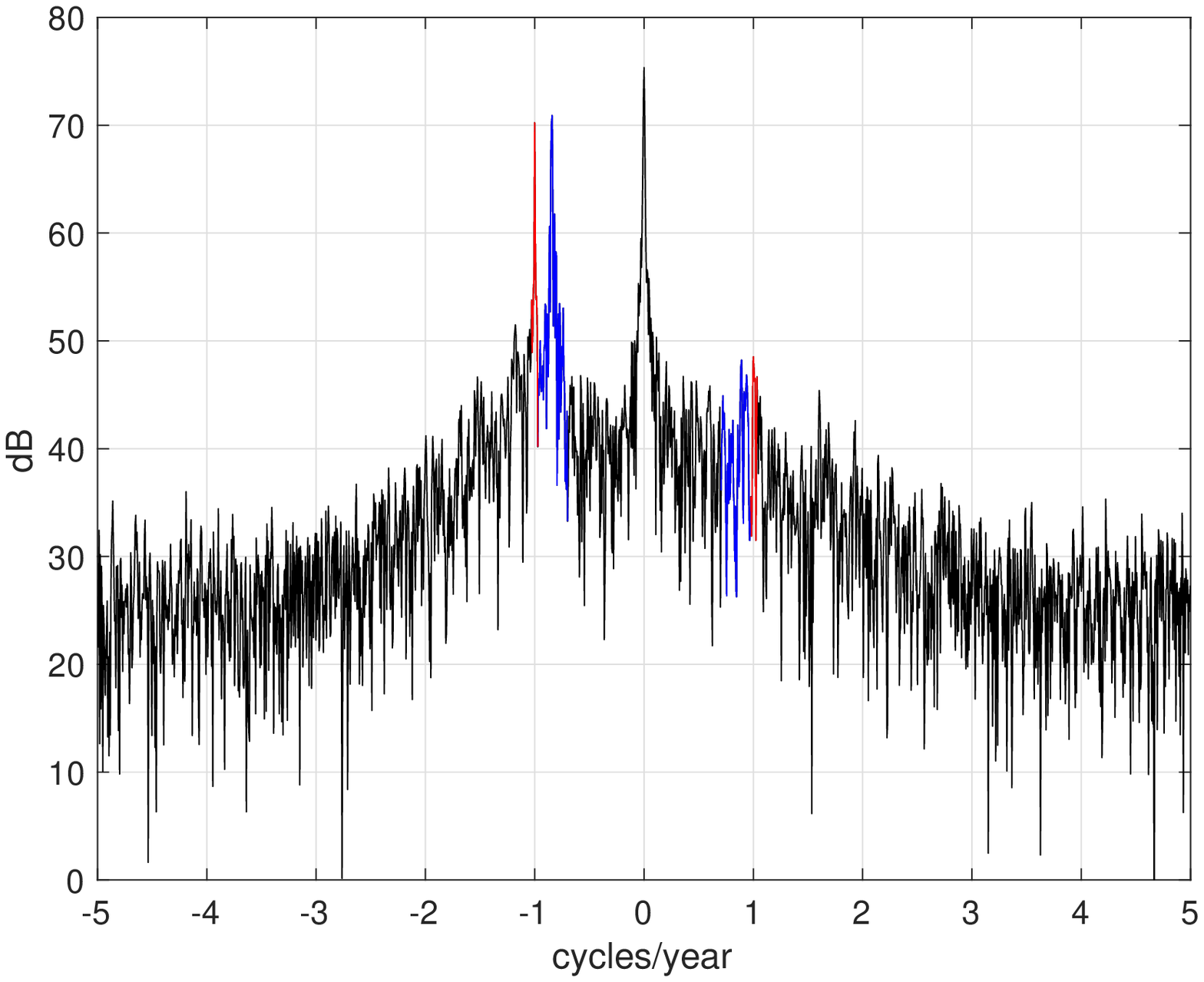}
\caption{(Left) Earth's polar motion from Dec 1845 to Oct 2021, measured in regular intervals of 0.1 years. (Right) The periodogram of Earth's polar motion of Fig.~\ref{fig2} when represented as a complex-valued time series. The red band of frequencies corresponds to the annual oscillation, and the blue band to the Chandler wobble.}
\label{fig2}
\end{figure}

Polar motion measures the deviation of Earth's rotational axis relative to its crust. In the left panel of Fig.~\ref{fig2} we plot Earth's polar motion from 1845 to 2021 in orthogonal $x$ and $y$ directions, as measured in milliarcseconds (mas), where 100mas corresponds to a deviation of approximately 3 metres at the Earth's surface. This data is publicly available from the International Earth Rotation and Reference Systems Service Earth Orientation products\footnote{www.iers.org/IERS/EN/DataProducts/EarthOrientationData/eop}. We observe a slow drift in the time series, especially in the $y$ axis, coupled with clear oscillatory motion. We are motivated to study this dataset in particular because~\cite{arato1962evaluation} also studied Earth's polar motion when proposing the complex OU process. Here we can make use of over 50 years' worth of new data to provide updated parameter estimates, and test for the presence of ellipticity using our elliptical OU process of~\eqref{eq:WILCOU}.

In the right panel of Fig.~\ref{fig2} we plot the periodogram of the complex-valued time series $z(t)=x(t)+iy(t)$. We detect three clear peaks in the periodogram. The largest at frequency zero is due to the drift. The smallest, at (negative) one cycle per year, is the annual oscillation. The third, at approximately -0.84 cycles per year is the Chandler wobble, discovered by astronomer Seth Carlo Chandler in 1891.

We will study the properties of both oscillations using the elliptical OU of~\eqref{eq:WILCOU}. To do this we cannot simply look at the precise values and locations of the peaks in the spectral density---we also need to consider frequencies in the {\em vicinity} of the peaks, such that we can estimate the damping parameter $\alpha_1$ of the oscillations. We have marked in blue and red in the right panel of Fig.~\ref{fig2} (respectively) the frequencies we will use to model the Chandler and Earth wobble oscillations respectively. Specifically, the Chandler Wobble is considered in the range -0.97 to -0.70 cycles per year, and the annual oscillation in the range -1.03 to -0.97 cycles per year. We have also marked the corresponding positive frequencies, which will contain some elevated power if this component of the time series has elliptical structure.

\begin{figure}[h]
\centering
\includegraphics[width=3in]{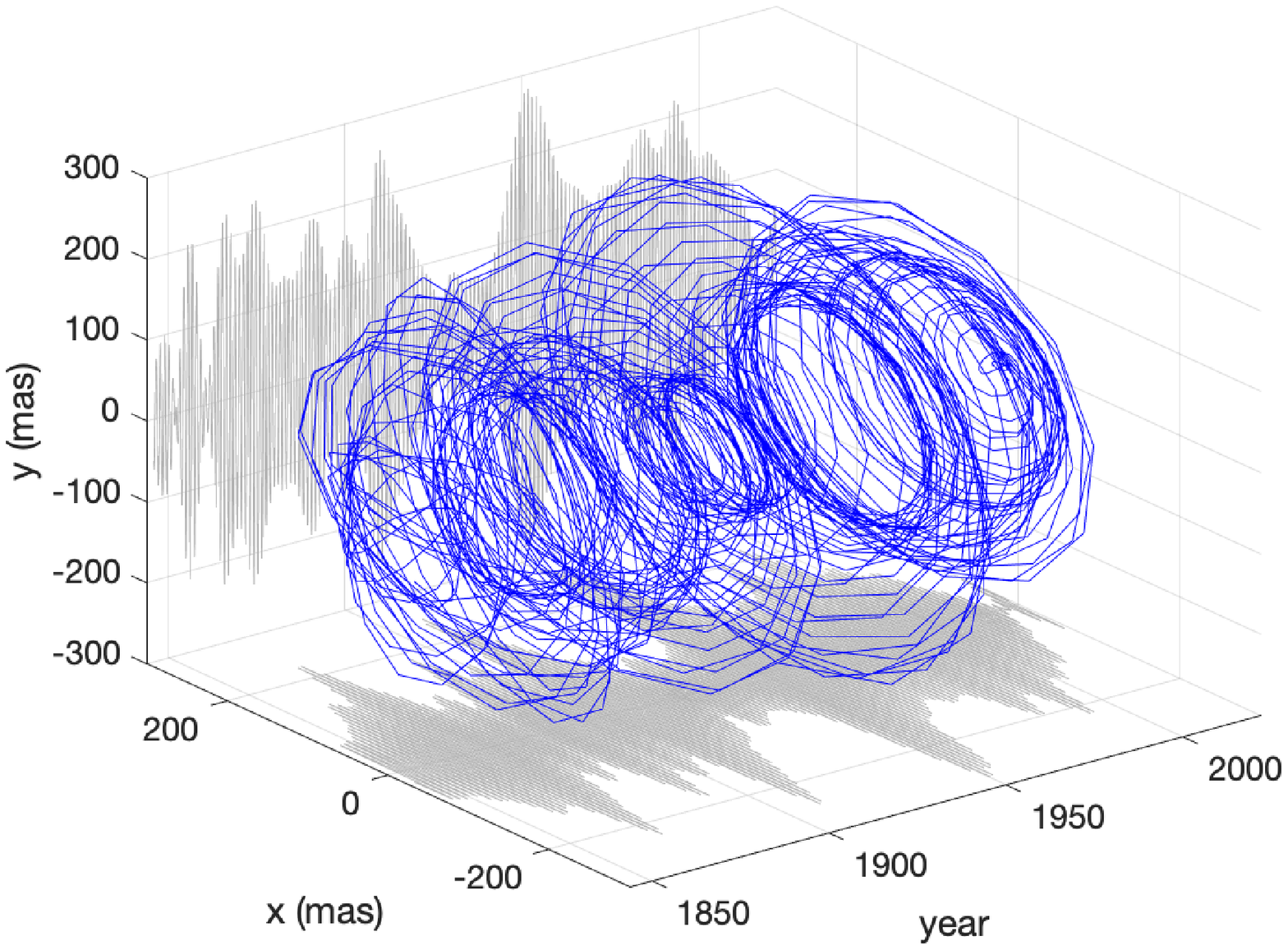} \hspace{1cm}
\includegraphics[width=3in]{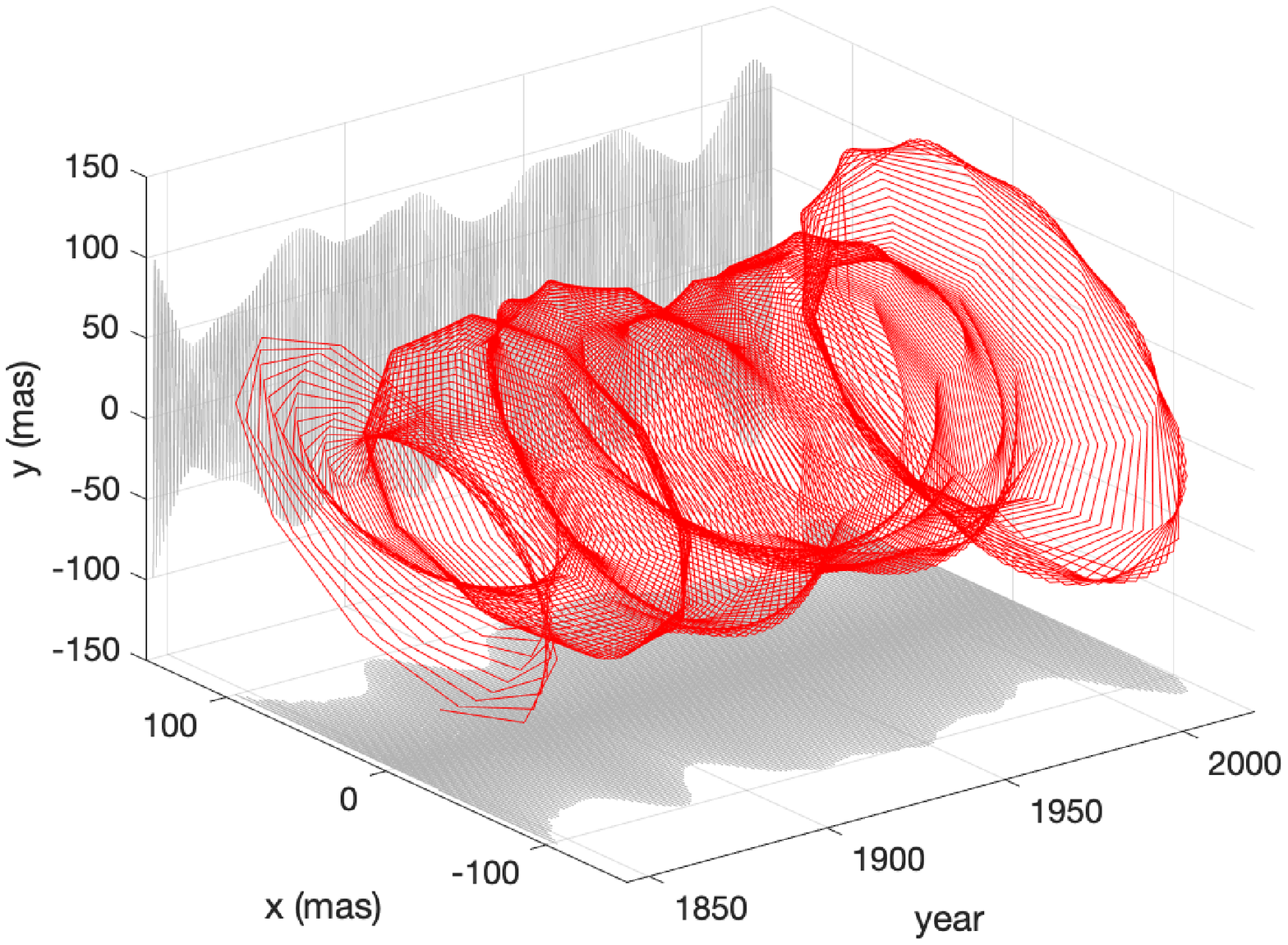}
\caption{(Left) The Chandler wobble over time, which has been bandpass filtered from Fig.~\ref{fig2} with a boxcar filter using the blue frequencies highlighted in the periodogram. (Right) The annual oscillation which has been bandpass filtered from Fig.~\ref{fig2} with a boxcar filter using the red frequencies highlighted in the periodogram.\vspace{2mm}}
\label{fig5}
\end{figure}

For visualisation, we bandpass filter the polar motion time series with boxcar filters and display the resulting time series in Fig.~\ref{fig5}. The left panel uses the blue frequencies from Fig.~\ref{fig2} and therefore corresponds to the Chandler wobble motion, and the right panel uses the red frequencies and therefore corresponds to the annual oscillation. The presence of damping can be seen in both oscillations, especially the Chandler wobble, which motivated the construction of the complex OU by Arat\'o {\em et al.} in~\cite{arato1962evaluation}. No clear ellipticity is seen in the Chandler wobble oscillations, but there appears to be some present in the annual oscillations. We will study this in more detail by fitting complex and elliptical OU processes to this data. Note that the filtering procedure performed here (and choice of filter) is purely for visualisation, and does not impact the parameter estimation, as the periodogram of the unfiltered time series is used in the Whittle likelihood.

\begin{figure}
\centering
\includegraphics[width=2.9in]{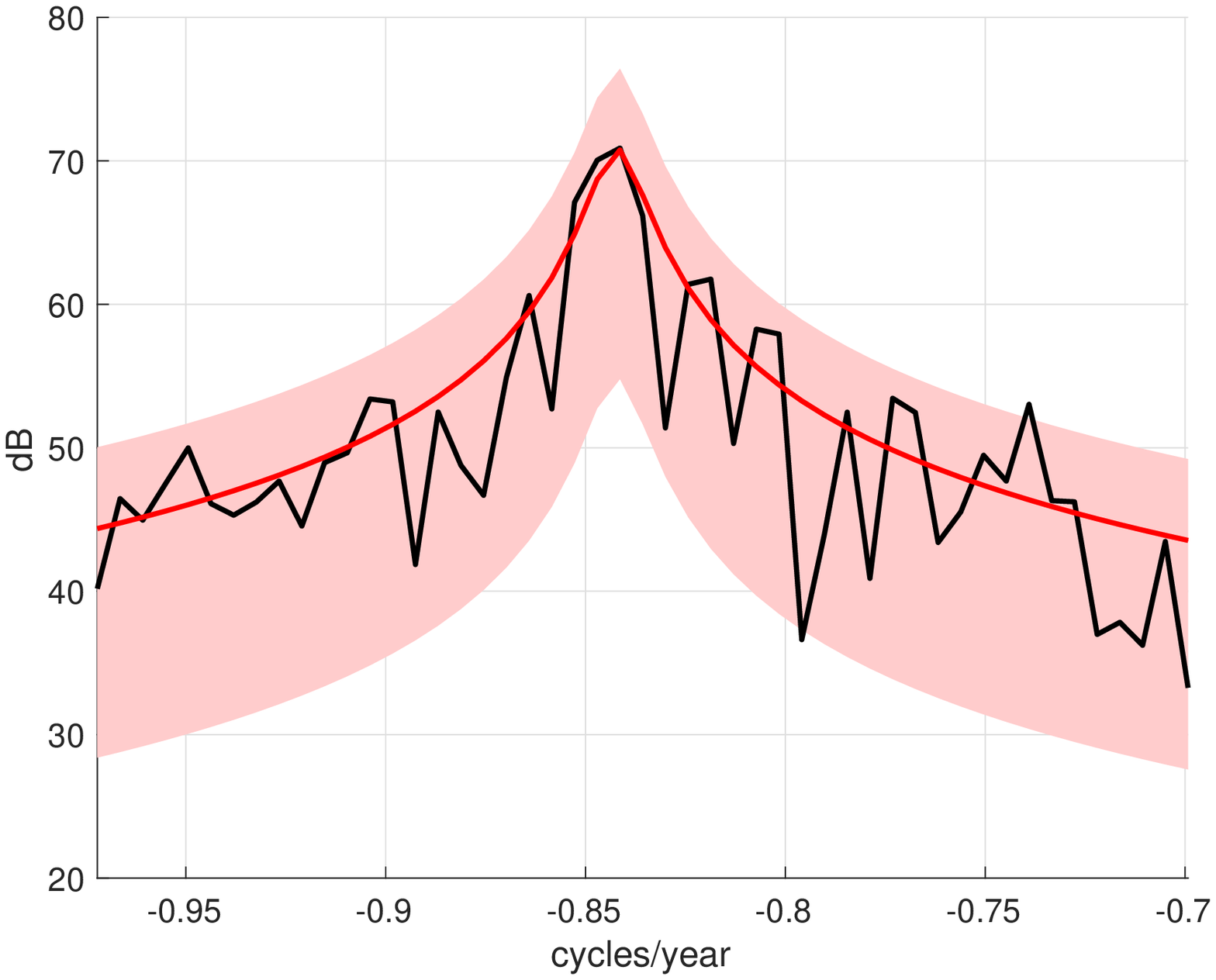} \hspace{1cm}
\includegraphics[width=2.9in]{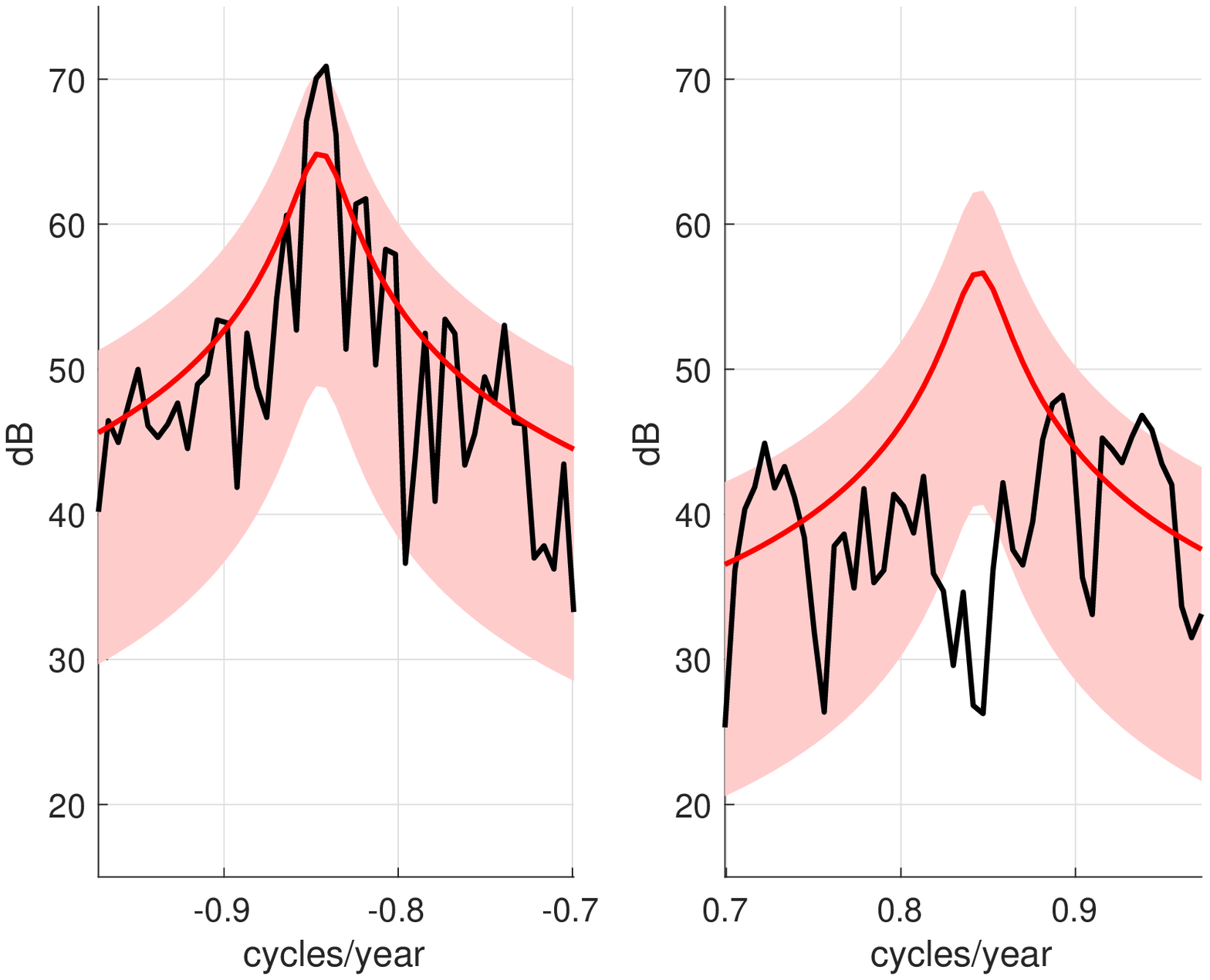}
\caption{Semi-parametric Whittle fits of modelled spectra (red) to the observed periodogram (black) of Earth's polar motion. In the left panel we fit the complex OU spectrum of~\eqref{eq:psd} in the frequency interval of -0.96 to -0.73 cycles per year which captures the Chandler wobble. In the right two panels we fit the elliptical OU spectrum of~\eqref{eq:psd2} in the frequency intervals of -0.96 to -0.73 and 0.73 to 0.96 cycles per year. In all panels 95\% pointwise confidence intervals of the power spectral density are in red shading.}
\label{fig9}
\end{figure}

First, we consider the Chandler Wobble over negative frequencies only and fit the complex OU of~\cite{arato1962evaluation}, which corresponds to the elliptical OU of~\eqref{eq:WILCOU} when $\alpha_2=\beta_2=r=0$. We fit the parameters using the semi-parametric Whittle likelihood of \eqref{eq:Whittle2}, using only negative frequencies in the interval $\omega \in [-0.97,-0.7]$. The fit of the power spectral density of the complex OU in~\eqref{eq:psd} to the periodogram is displayed in the left panel of Fig.~\ref{fig9}. Although the periodogram is variable, it lies within the 95\% pointwise confidence intervals of the modelled power spectral density almost everywhere. Confidence intervals are estimated using the asymptotic exponential distribution of the periodogram. The estimated parameters (to 3 significant figures) are $\alpha_1 = 0.0389$ (in units of years$^{-1}$), $\beta_1 = -0.842$ (cycles per year) and $\sigma^2=183$.  Using the bootstrap procedure described in Section~\ref{ss:bootstrap} we obtain 95\% confidence intervals of $[0.0167,0.102]$ for $\alpha_1$, $[-0.847,-0.835]$ for $\beta_1$, and $[119,266]$ for $\sigma^2$. In this section we increase the number of bootstrap replicates to 10,000 such that the approximated confidence intervals converge to the resolution provided.

Arat\'o {\em et al.} \cite{arato1962evaluation}, in their 1962 analysis, found $\alpha_1=0.06$ and $\beta_1 = -0.839$, but the 95\% confidence range for $\alpha_1$ was found to be $[0.008,0.13]$ which is overall consistent with our estimates and confidence intervals, but we find a slightly lower damping parameter after utilising over 50 years' worth of new data. However, information about the damping parameter lies over very few frequencies, and is therefore a challenging parameter to estimate, as also observed by~\cite{arato1962evaluation} and in our simulation studies of Section~\ref{ss:sim}. In other literature, Brillinger \cite{brillinger1973empirical} also uses a complex OU process like Arat\'o {\em et al.}, but makes some seasonal corrections, and also finds $\alpha_1=0.06$ cycles per year with a 95\% confidence range of $[0.006 ,0.114]$. More broadly, there still remains an active research debate on the rate of damping of the Chandler wobble~\cite{vondrak2017new}, where a variety of geophysical models have been employed to measure this, but a more detailed comparison with this literature is beyond the scope of this paper.

We next fit the elliptical OU process to the Chandler wobble over negative and positive frequencies $\omega \in \pm[0.7,0.97]$, using \eqref{eq:Whittle2}. The fits are displayed in the right panel of Fig.~\ref{fig9}. Clearly the fit to positive frequencies is poor, with no observable peak in the periodogram at the expected oscillation frequency of the Chandler wobble, and several values lying outside of the 95\% pointwise confidence intervals of the modelled spectrum. Fitting parameters using \eqref{eq:Whittle1} which incorporates the complementary spectrum did not improve the fits. Overall, there is insufficient evidence to support the presence of a positive-frequency peak in the Chandler wobble corresponding to an elliptical oscillatory motion.
This is consistent with the literature where the Chandler wobble motion has been described as ``{\em quasi}-circular" with a very low eccentricity in the range $[0.1, 0.23]$ in~\cite{hopfner2003chandler}. In our case, the periodogram of the time series is too variable and contaminated by other artefacts at positive frequencies, so our model is unable to detect this low eccentricity in the Chandler wobble oscillation.

\begin{figure}
\centering
\includegraphics[width=3in]{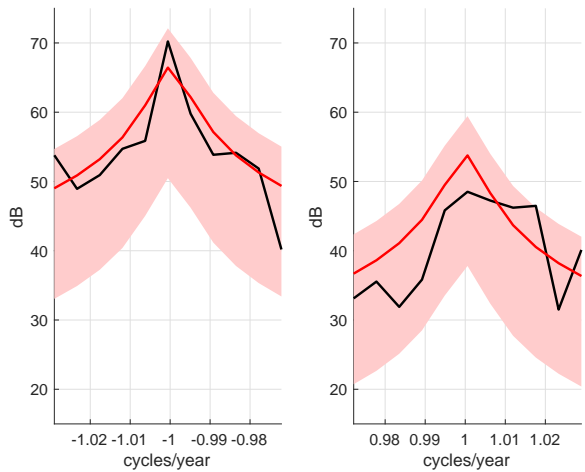}
\caption{Semi-parametric Whittle fits of the elliptical OU spectral density of~\eqref{eq:psd2} (red) to the observed periodogram (black) of Earth's polar motion. Fits performed over frequency intervals of -1.035 to -0.965 and 0.965 to 1.035 cycles per year, thus capturing the annual oscillation. 95\% pointwise confidence intervals of the power spectral density are in red shading.}
\label{fig11}
\end{figure}

Finally, we fit the elliptical OU process to the annual oscillation over negative and positive frequencies $\omega \in \pm[0.97,1.03]$ using \eqref{eq:Whittle2}, where we fix the peak frequency $\beta$ to be 1 cycle per year, leaving just three parameters $\{\alpha,\rho,A^2\}$ to be estimated. This simplification was required to make the optimisation feasible as our effective sample size is only 44 (11 positive and 11 negative Fourier frequencies each containing an amplitude and phase) due to the narrow modelling interval. Inference using \eqref{eq:Whittle1} was not possible in this example, as parameters converged to boundary values.
The fitted spectra using \eqref{eq:Whittle2} are displayed in Fig.~\ref{fig11}, and this time the model is a good fit, and the periodogram comfortably lies withing the 95\% confidence interval bands at all modelled frequencies. The estimated parameters of the elliptical OU and their 95\% confidence intervals are given in Table~\ref{Tab4}. Due to the narrowband nature of the annual signal, the use of this model and the estimated parameter values should be interpreted with some caution, as reflected by the wide confidence intervals. The orientation parameter $\psi$ was estimated non-parametrically using~\eqref{eq:orient} to be $\psi=0.125$, and the eccentricity of the annual oscillation was estimated to be $\varepsilon=\sqrt{1-\rho^4}=0.782$ (with a 95\% confidence interval of [0.639,0.915]). This is in broad agreement but somewhat different from~\cite{hopfner2003chandler} who discover a ``significantly
elliptic annual motion" in the range $[0.26,0.49]$. For comparison, a simple non-parametric estimate using (see~\cite{sykulski2016widely})
\[
\hat\varepsilon = \frac{2\sqrt{|J_Z(\omega_{\max})J_Z(-\omega_{\max})|}}{|J_Z(\omega_{\max})|+|J_Z(-\omega_{\max})|},
\]
yields $\hat\varepsilon=0.530$. The higher values of eccentricity we estimate as compared with~\cite{hopfner2003chandler} are likely due to their approach of time-windowing into small intervals, versus our approach of considering the entire time series as one stochastic process. Again, a more detailed analysis is beyond the scope of this paper, however our example here serves as a simple proof-of-concept of the potential applications of our novel elliptical OU SDE model.

\begin{table}[h]
\begin{center}
\caption{\label{Tab4}Elliptical OU parameter estimates and 95\% confidence interval limits}
\begin{tabular}{|c|ccccc|} \hline
& $\alpha_1$ & $\beta_1$ & $\alpha_2$ & $\beta_2$ & $\sigma^2$ \\ \hline
estimate & 0.0245 & -1.11 & 0.122 & 0.476 & 28.4 \\
lower limit & 0.0091 & -1.44 & 0.066 & 0.258 & 11.4 \\
upper limit & 0.0929 & -1.03 & 0.257 & 1.00 & 74.7 \\ \hline
\end{tabular}
\end{center}
\end{table}

\section{Discussion and Conclusion}\label{S:Conc}
Oscillations are key features of natural and human-made phenomena. Often we observe linked oscillations that map out the same periodic phenomenon. For deterministic phenomena, such have been studied in \cite{lilly2011analysis,olhede2013modulated}, and for stochastic phenomena in \cite{sykulski2016lagrangian,sykulski2016widely}. Continuous-time time series that are improper are, as we have shown, challenging to describe but possess interpretable multidimensional dynamics \cite{lilly2011analysis}. The aim of this paper has been to introduce a structured form of multivariate dependence so that stochastic elliptical trajectories are mapped out, just like single oscillations can be conceptualised as mapping out circles. Complex-valued models, such as the elliptical OU process, provide rich structural information as we can recover the geometric features of the ellipse directly from the observations and estimated parameters.

Multivariate stochastic processes have been the focus of intensive research in the last decade \cite{barigozzi2018simultaneous,chang2018principal, che2018recurrent,hallac2017toeplitz,nieto2016common}. There is much advantage to modelling underlying geometry in time series \cite{olhede2005local}, but that viewpoint exactly corresponds as to how the underlying structure in the observations evolves over time. Oscillations are natural as a modelling starting point when studying stationary phenomena. The multivariate generalisation of an oscillation is an observed trajectory from an ellipse \cite{lilly2011analysis}. This puts an emphasis on the classes of models starting from oscillations, broadening to partially observed trajectories on the ellipse.

A number of questions remain unresolved. Our generalisation of the OU model is just one example of a statistical model of temporal structure. The differential equation linkage has been discussed further for other applications including random fields by \cite{lindgren2011explicit}. We envision that similar extensions could be done to their model classes. This would build on the non-parametric statistical work of~\cite{walden2013rotary}. Furthermore, we can seek similar extensions to trivariate and multivariate time series, building stochastic analogues to the deterministic approaches taken in~\cite{lilly2011analysis}.

Finally, inspired by the works of \cite{arato1962evaluation,baran2018d,brillinger1973empirical}, the applicability of the elliptical OU process has been demonstrated by the analysis of Earth's polar motion and the Chandler wobble. In the future, such analysis can be repeated on other planets such as Mars \cite{van2000chandler}, especially as richer datasets become available making such studies more feasible. The challenges of real data examples will stress test our model, and show us what new features and geometrical structures require incorporating into the model framework.

All results, figures and tables in this paper can be exactly reproduced using the MATLAB code available for free download at \url{https://github.com/AdamSykulski/EllipticalOU}.

\section*{Appendix A: Relationship between the OU and AR(1) Processes}
Consider the widely linear complex autoregressive process of~\cite{sykulski2016widely} given by
\begin{equation}
Z(t) = \lambda e^{i\zeta} Z(t-1) + \gamma e^{i\phi} Z^*(t-1) + \epsilon_t,
\label{eq:WILCAR}
\end{equation}
with noise variance $\sigma_{AR}^2$ and pseudo-variance $r_{AR}$. Let us now contrast this with the elliptical OU process of~\eqref{eq:WILCOU}. In the simple (proper) case of $\gamma = \alpha_2 = \beta_2 = r = r_{AR} = 0$ then a sampled complex OU (at intervals $\Delta$) is like a complex AR(1) where
\begin{align}
 \lambda &= e^{-\alpha_1\Delta},
 \label{eq:appa1}
\\ \nonumber
 \sigma_{AR}^2 &= \sigma^2\frac{(1-e^{-2\alpha_1\Delta})}{2\alpha_1},
\\ \nonumber
 \zeta &= \beta_1,
\end{align}
thus providing a simple mapping between the processes. These relationships can be derived by considering an Euler-Maruyama expansion of the OU:
\[
z(t+1/x) \approx (1-\alpha_1/x+i\beta_1/x)z(t) + \sqrt{A^2/x}B,
\]
where $x$ is large and $B$ is a draw from a $\mathcal{N}(0,1)$ such that repeating $x\Delta$ times we have
\[
z(t+\Delta) (1-\alpha_1/x+i\beta_1/x)^{x\Delta}z(t) + \sqrt{A^2/x} \sum_{k=0}^{x\Delta-1}(1-\alpha_1/x)^{k}B,
\]
and then taking $x\rightarrow\infty$ we get the relationships above.

In the general case $\gamma \neq \alpha_2 \neq \beta_2 \neq r \neq r_{AR} \neq 0$ then the Euler-Maruyama expansion becomes
\begin{align*}
z\left(t+\frac{1}{x}\right) \approx & \left(1-\frac{\alpha_1}{x}+\frac{i\beta_1}{x}\right)z(t) - \left(\frac{\alpha_2}{x}-\frac{i\beta_2}{x}\right) z^*(t) + \sqrt{\frac{1}{x}}B,
\end{align*}
where $B$ is a draw from $\mathcal{CN}(0,\sigma^2,r)$. Then
repeating $x\Delta$ times and taking $x\rightarrow\infty$ we observe that
\begin{align*}
\lambda e^{i\zeta} = & \lim_{x\rightarrow\infty} \sum_{k=0}^{x\Delta/2} \left(1-\frac{\alpha_1}{x}+\frac{i\beta_1}{x}\right)^{x\Delta-2k} \left(\frac{\alpha_2}{x}-\frac{i\beta_2}{x}\right)^{2k} \binom{x\Delta}{2k},
\\
\gamma e^{i\phi} = & \lim_{x\rightarrow\infty} \sum_{j=1}^{x\Delta/2} \left(1-\frac{\alpha_1}{x}+\frac{i\beta_1}{x}\right)^{x\Delta-2j+1} \left(\frac{\alpha_2}{x}-\frac{i\beta_2}{x}\right)^{2j-1} \binom{x\Delta}{2j-1},
\end{align*}
which have no clear analytical solutions, such that we can observe the nontrivial mapping between the processes in the widely linear case.

\section*{Appendix B: Power spectral density derivation}
To derive the power spectral density of the elliptical OU process, we start from the power spectral density of the complex OU in~\eqref{eq:psd} and convert to Cartesian forms using the relationships given in~\cite{sykulski2017frequency}
\begin{align*}
    S_{\tilde x}(\omega)&  = \frac{1}{4}\{S_{\tilde z}(\omega) + S_{\tilde z}(-\omega)\}+\frac{1}{2}\mathcal{R}\{R_{\tilde z}(\omega)\}, \nonumber \\
    S_{\tilde y}(\omega)&  = \frac{1}{4}\{S_{\tilde z}(\omega) + S_{\tilde z}(-\omega)\}-\frac{1}{2}\mathcal{R}\{R_{\tilde z}(\omega)\}, \nonumber \\
    S_{\tilde x\tilde y}(\omega)&  = 
    \frac{1}{2}\mathcal{I}\{R_{\tilde z}(\omega)\} +
    \frac{i}{4}\{S_{\tilde z}(\omega) - S_{\tilde z}(-\omega)\},
\end{align*}
where $S_{\tilde x\tilde y}(\omega)$ is the cross-spectral density between $\tilde x(t)$ and $\tilde y(t)$, and $\mathcal{R}\{\cdot\}$ and $\mathcal{I}\{\cdot\}$ denote the real and imaginary part respectively. Substituting in~\eqref{eq:psd}, and using that $R_{\tilde z}(\omega)=0$ as the complex OU is a proper process, we obtain
\begin{align}
    S_{\tilde x}(\omega) &= \frac{A^2}{4}\left\{ \frac{1}{\alpha^2+(\omega-\beta)^2} + \frac{1}{\alpha^2+(\omega+\beta)^2} \right\}, \label{app1}\\
    S_{\tilde y}(\omega)  &=\frac{A^2}{4}\left\{ \frac{1}{\alpha^2+(\omega-\beta)^2} + \frac{1}{\alpha^2+(\omega+\beta)^2} \right\},  \\
    S_{\tilde x\tilde y}(\omega)&  = \frac{iA^2}{4}\left\{ \frac{1}{\alpha^2+(\omega-\beta)^2} - \frac{1}{\alpha^2+(\omega+\beta)^2} \right\}. \label{app2}
\end{align}
Note that $S_{\tilde x}(\omega)=S_{\tilde y}(\omega)$. Next we find the power spectral densities of the elliptically transformed bivariate OU process of~\eqref{eq:bvou3}. First we note by expanding~\eqref{eq:bvou2} that
\[
    x(t) = \frac{1}{\rho}\tilde x(t)\cos\psi -\rho \tilde y(t)\sin\psi, \quad
    y(t) = \rho \tilde y(t)\cos\psi + \frac{1}{\rho}\tilde x(t) \sin\psi.
\]
This clarifies the geometric interpretation of $P$ and $Q$ in~\eqref{eq:bvou2}. It then follows that
\begin{align}
\label{app3}
    S_{x}(\omega) = & \frac{\cos^2 \psi}{\rho^2} S_{\tilde x}(\omega) + \rho^2 \sin^2\psi S_{\tilde y}(\omega) -\cos\psi\sin\psi S_{\tilde x\tilde y}(\omega)  - \cos\psi\sin\psi S^*_{\tilde x\tilde y}(\omega), \\
    S_{y}(\omega) = & \frac{\sin^2 \psi}{\rho^2} S_{\tilde y}(\omega) + \rho^2 \cos^2\psi S_{\tilde x}(\omega) + \cos\psi\sin\psi S_{\tilde x\tilde y}(\omega)  + \cos\psi\sin\psi S^*_{\tilde x\tilde y}(\omega), \\
    \label{app4}
    S_{xy}(\omega)  = & \frac{\cos\psi\sin\psi}{\rho^2}S_{\tilde x}(\omega) - \rho^2\cos\psi\sin\psi S_{\tilde y}(\omega) + \cos^2\psi S_{\tilde x\tilde y}(\omega) - \sin^2\psi S^*_{\tilde x\tilde y}(\omega),
\end{align}
where we have used that $S_{\tilde y\tilde x}(\omega)=S^*_{\tilde x\tilde y}(\omega)$.
Substituting \eqref{app1}--\eqref{app2} into \eqref{app3}--\eqref{app4} we obtain
\begin{align}
\label{app5}
    S_{x}(\omega) = & \frac{A^2}{4}\left(\frac{\cos^2 \psi}{\rho^2}+ \rho^2\sin^2\psi \right)  \left\{ \frac{1}{\alpha^2+(\omega-\beta)^2} + \frac{1}{\alpha^2+(\omega+\beta)^2} \right\}, \\
    S_{y}(\omega) = & \frac{A^2}{4}\left(\frac{\sin^2 \psi}{\rho^2}+ \rho^2\cos^2\psi \right) \left\{ \frac{1}{\alpha^2+(\omega-\beta)^2} + \frac{1}{\alpha^2+(\omega+\beta)^2} \right\}, \\
    \label{app6} 
    S_{xy}(\omega) = & \frac{A^2}{4}\left(\frac{\cos\psi\sin\psi}{\rho^2} - \rho^2\cos\psi\sin\psi \right) \left\{ \frac{1}{\alpha^2+(\omega-\beta)^2} + \frac{1}{\alpha^2+(\omega+\beta)^2} \right\} \\ \nonumber &+ \frac{iA^2}{4}\left\{ \frac{1}{\alpha^2+(\omega-\beta)^2} - \frac{1}{\alpha^2+(\omega+\beta)^2} \right\}.
\end{align}
We then convert back to complex using the relationships given in~\cite{sykulski2017frequency}
\begin{align}
    S_z(\omega) & = S_x(\omega) + S_y(\omega) + 2\mathcal{I}\{S_{xy}(\omega)\}, \label{app7} \\
    R_z(\omega) & = S_x(\omega) - S_y(\omega) + 2i\mathcal{R}\{S_{xy}(\omega)\}. \label{app8}
\end{align}
Substituting \eqref{app5}--\eqref{app6} into \eqref{app7}--\eqref{app8} we obtain
\begin{align*}
    S_z(\omega) =& \frac{A^2}{4}\left(\frac{1}{\rho^2}+ \rho^2 \right) \left\{ \frac{1}{\alpha^2+(\omega-\beta)^2} + \frac{1}{\alpha^2+(\omega+\beta)^2} \right\} + \frac{A^2}{2} \left\{ \frac{1}{\alpha^2+(\omega-\beta)^2} - \frac{1}{\alpha^2+(\omega+\beta)^2} \right\}, \\
    R_z(\omega) =&\frac{A^2}{4}\left(\frac{1}{\rho^2}- \rho^2 \right)(\cos^2\psi-\sin^2\psi) \left\{ \frac{1}{\alpha^2+(\omega-\beta)^2} + \frac{1}{\alpha^2+(\omega+\beta)^2} \right\} \\ & + \frac{iA^2}{2}\left(\frac{1}{\rho^2}- \rho^2 \right)\cos\psi\sin\psi  \left\{ \frac{1}{\alpha^2+(\omega-\beta)^2} + \frac{1}{\alpha^2+(\omega+\beta)^2} \right\},
\end{align*}
which simplify to the forms given in~\eqref{eq:psd2} and \eqref{eq:psd3}.

\section*{Appendix C: Non-parametric estimation of the orientation}
Here we derive the form of the non-parametric estimate given in~\eqref{eq:orient}. From~\eqref{eq:psd3} we have that $\arg\{R_z(\omega)\}=2\psi$, from which we can form the direct estimate
\begin{align*}
    2\hat\psi = & \arg\{\hat R_Z(\omega)\} = \arg\{J_Z(\omega)J^\ast_{Z^\ast}(\omega)\} \\
    = & \arg\{J_Z(\omega)\} + \arg\{J^*_{Z^\ast}(\omega)\} \\
    = & \arg\{J_Z(\omega)\} + \arg\{J_Z(-\omega)\},
\end{align*}
where we have used the cross-periodogram estimator given by $\hat R_Z(\omega) =J_Z(\omega)J^\ast_{Z^\ast}(\omega)$ to estimate $ R_z(\omega)$. We evaluate the above expression at $\omega = \omega_{\max}$ where $\omega_{\max}$ is the known/estimated location of the peak in the power spectral density of the elliptical OU process, which in the application to the annual oscillation of Earth's polar motion this peak occurs at $\omega_{\max} = 1$ cycle per year.

\end{document}